\documentclass[aps,prd,nofootinbib,superscriptaddress,preprintnumbers,amsmath,amssymb,latexsym,array,enumerate,letterpaper,twocolumn]{revtex4}
\usepackage{amssymb}
\usepackage{amsmath}
\usepackage{epsfig}
\usepackage{hyperref}
\usepackage{breakurl}
\usepackage{xcolor}

\makeatletter
\def\simgt{\mathrel{\lower2.5pt\vbox{\lineskip=0pt\baselineskip=0pt
           \hbox{$>$}\hbox{$\sim$}}}}
\def\simlt{\mathrel{\lower2.5pt\vbox{\lineskip=0pt\baselineskip=0pt
           \hbox{$<$}\hbox{$\sim$}}}}
\makeatother

\newcommand{\nc}{\newcommand}
\nc{\beq}{\begin{equation}}
\nc{\eeq}{\end{equation}}
\nc{\barray}{\begin{eqnarray}}
\nc{\earray}{\end{eqnarray}}
\nc{\barrayn}{\begin{eqnarray*}}
\nc{\earrayn}{\end{eqnarray*}}
\nc{\bcenter}{\begin{center}}
\nc{\ecenter}{\end{center}}
\nc{\mc}{\mathcal}
\nc{\er}[1]{(\ref{eq:#1})}
\nc{\onehalf}{\frac{1}{2}} 
\nc{\partialbar}{\bar{\partial}}
\nc{\psit}{\widetilde{\psi}}
\nc{\Tr}{\mbox{Tr}}
\nc{\hc}{\mbox{H.c.}}
\nc{\ev}{\;\mathrm{eV}}
\nc{\mev}{\;\mathrm{MeV}}
\nc{\gev}{\;\mathrm{GeV}}
\nc{\kev}{\;\mathrm{keV}}
\nc{\tev}{\;\mathrm{TeV}}

\def\chii0{\chi_i^0}
\def\chij0{\chi_j^0}

\newcommand{\gsim}{\lower.7ex\hbox{$\;\stackrel{\textstyle>}{\sim}\;$}}
\newcommand{\lsim}{\lower.7ex\hbox{$\;\stackrel{\textstyle<}{\sim}\;$}}
\nc{\ttbar}{t\bar t}

\def\beq{\begin{equation}}
\def\eeq{\end{equation}}
\def\bea{\begin{eqnarray}}
\def\eea{\end{eqnarray}}

\usepackage{comment}

\newcommand{\fref}[1]{Fig.~\ref{fig.#1}}
\newcommand{\eref}[1]{Eq.~(\ref{eq.#1})}

\newcommand{\erefs}[2]{Eqs.~(\ref{eq.#1})$\,-\,$(\ref{eq.#2})}
\newcommand{\aref}[1]{Appendix~\ref{a.#1}}

\newcommand{\cref}[1]{Chapter~\ref{c.#1}}
\newcommand{\tref}[1]{Table~\ref{t.#1}}

\def\beq{\begin{equation}}
\def\eeq{\end{equation}}
\def\bea{\begin{eqnarray}}
\def\eea{\end{eqnarray}}

\graphicspath{{plots/}}

\begin{document}

\widetext

\title{``Non-Local'' Effects from Boosted Dark Matter in Indirect Detection}

\author{Kaustubh Agashe}
\affiliation{Maryland Center for Fundamental Physics, Department of Physics, University of Maryland, College Park, MD 20742-4111 USA}

\author{Steven J. Clark}
\affiliation{Brown Theoretical Physics Center and Department of Physics, Brown University, Providence, RI 02912-1843, USA}

\author{Bhaskar Dutta}
\affiliation{Mitchell Institute for Fundamental Physics and Astronomy, Department of Physics and Astronomy, Texas A\&M University, College Station, TX 77843, USA}

\author{Yuhsin Tsai}
\affiliation{Maryland Center for Fundamental Physics, Department of Physics, University of Maryland, College Park, MD 20742-4111 USA}
\affiliation{Department of Physics, University of Notre Dame, IN 46556, USA}

\preprint{UMD-PP-020-1,\,MI-TH-2016}

\begin{abstract}
Indirect dark matter (DM) detection typically involves the observation of standard model (SM) particles emerging from DM annihilation/decay inside regions of high dark matter concentration. We consider an annihilation scenario in which this reaction has to be initiated by one of the DMs involved being {\em boosted} while the other is an ambient non-relativistic particle. This ``trigger" DM must be created, for example, in a previous annihilation or decay of a heavier component of DM. Remarkably, boosted DM annihilating into gamma-rays at a specific point in a galaxy could actually have traveled from its source at another point in the same galaxy or even from another galaxy. Such a ``non-local'' behavior leads to a non-trivial dependence of the resulting photon signal on the galactic halo parameters, such as DM density and core size, encoded in the so-called ``astrophysical" $J$-factor. These non-local $J$-factors are strikingly different than the usual scenario. A distinctive aspect of this model is that the signal from dwarf galaxies relative to the Milky Way tends to be suppressed from the typical value to various degrees depending on their characteristics. This feature can thus potentially alleviate the mild tension between the DM annihilation explanation of the observed excess of $\sim$ GeV photons from the Milky Way's galactic center vs.~the apparent non-observation of the corresponding signal from dwarf galaxies.
\end{abstract}
 
\maketitle

\section{Introduction}

The search of dark matter (DM) annihilation or decay in experiments designed primarily to detect cosmic-ray particles (such as positrons and antiprotons) and gamma-rays, despite being called \emph{indirect} detection of DM, can provide direct information on many properties of DM particles inside galactic halos. For instance, the morphology of the signal shows the DM distribution inside galaxies, and the signal's energy and flux indicate the mass and the interaction strength of DM particles, respectively. Using a novel indirect DM detection scenario, we will illustrate in this work that a comparison of signals from {\em different} DM halos may even allow us to identify additional details of the generating process.

Over the past few years, several anomalies in astrophysical signatures have provided strong motivations to study such signals from DM models. Among the different searches, the Fermi-LAT experiment~\cite{Atwood_2009} produced a gamma-ray survey of the sky for $100~\mathrm{MeV}-100~\mathrm{GeV}$ scale photons for both the Milky Way (MW) and dwarf spheroidal galaxies (dSph). The experiment also observed an intriguing excess of gamma-rays from the MW center~\cite{TheFermi-LAT:2017vmf} (thus called the galactic center excess or GCE) that has the right morphology to be explained by DM physics~\cite{Hooper:2010mq}.
\footnote{It has also been proposed that unresolved gamma-ray point sources could account for the GCE, see for example~\cite{Abazajian:2010zy,Abazajian:2012pn,Lee:2015fea}. For a more recent discussion on this topic, see \cite{Leane:2019xiy,Chang:2019ars,Leane:2020nmi,Leane:2020pfc,Buschmann:2020adf}.}
As future experiments like e-ASTROGAM~\cite{DeAngelis:2017gra}, Gamma-400~\cite{Egorov:2020cmx}, and DAMPE~\cite{Duan:2017vqr} have been proposed to extend the energy coverage of the gamma-ray signal, we expect significant improvements in the observations of MW and dSph. We will therefore use the DM production of gamma-ray signal as an example to discuss how we can probe the dynamics of DM from an ensemble of such detections from different objects.

\begin{figure}
\includegraphics[width=0.95\columnwidth]{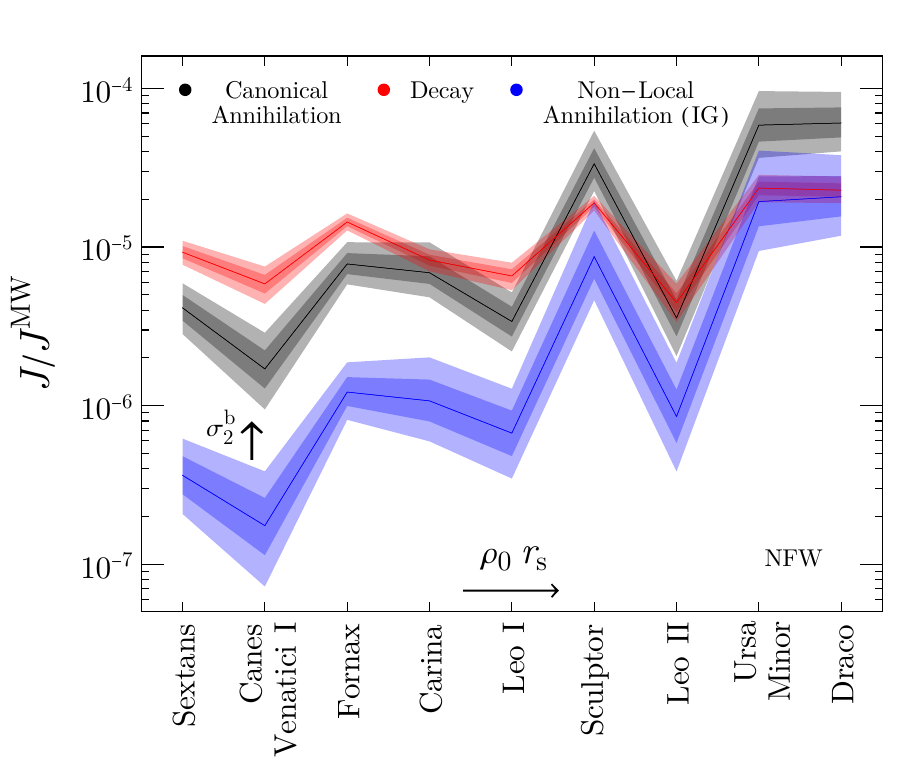}
\caption{The ratio of dSph $J$-factors to the MW's for various dark matter models assuming an NFW DM profile. The dSph are ordered by increasing values of $\rho_0\,r_s$ from left to right. We connect the results between dSph in order to better visualize the trend of galaxy-dependence. The width of the colored-bands at each galaxy represents the 1 and 2$\sigma$ uncertainties. dSph NFW profile parameters were obtained from~\cite{Pace:2018tin} and their central values are listed in \tref{galaxy_param} along with those for MW. As MW is used only as a reference here, we take $J^{\rm MW}$ as its central value. $\sigma_2^{\rm b}$ is the cross-section of the second annihilation process in the non-local model, see text for details. $\sigma_2^{\rm b}$ is chosen such that all galaxies have entered the non-local regime. The vertical arrow is a reminder that the non-local $J$-factor ratio can be larger for larger cross-sections. At its maximum, the $J$-factor ratio is indistinguishable from canonical annihilation. For the non-local annihilation, we only include intra-galactic contributions in this figure as noted by (IG). Here the region-of-interest was taken to be $\theta<0.5^\circ$ for the dSph and $\theta<45^\circ$ for MW. The line-of-sight integration extends out to 500 kpc.}
\label{fig.fingerprint}
\end{figure}

The differential photon flux ${\rm d}\Phi/{\rm d}E_{\gamma}$ arising from DM annihilation or decay in any astrophysical target for indirect DM detection 
is~\cite{Slatyer:2017sev}
\begin{equation}\label{eq.ndep1}
\frac{{\rm d}\Phi}{{\rm d}E_{\gamma}}=\frac{{\rm d}N}{{\rm d}E_{\gamma}}\begin{cases}\frac{\langle\sigma_{\rm ann} v\rangle}{8\pi\,m_{\chi}^2}\times J_{\rm ann} & \mbox{(annihilation)} \\ \frac{1}{4\pi\,m_{\chi}\,\tau_{\chi}}\times J_{\rm dec} & \mbox{(decay)}\end{cases}
\end{equation}
where the so-called $J$-factor encodes all the astrophysical contributions. ${\rm d}N/{\rm d}E_{\gamma}$ is the photon spectrum produced per annihilation or decay, $m_{\chi}$ is the DM particle mass, $\left\langle\sigma_{ \rm ann}v\right\rangle$ is the DM's thermally averaged annihilation rate with annihilation cross section $\sigma_{ \rm ann }$, and $\tau_{ \chi}$ is the DM lifetime. Everything except the $J$-factor is independent of the galactic environment and originates from the underlying particle physics. For instance, the $J$-factors for the ``canonical'' DM annihilation (by which
we mean the process of two ambient DM particles annihilating into SM particles) and decay that happen in a far away galaxy at a distance $d$ much larger than the galaxy's size are
\begin{equation}\label{eq.J}
J_{\rm ann}=d^{-2}\int {\rm d}V\rho^2(r)\,,\quad J_{\rm dec}=d^{-2}\int {\rm d}V\rho(r)\,,
\end{equation}
where $\rho$ is the DM density and the integral is performed over the galaxy's volume. The reader can consult \aref{Jfacor_alt_derv} for a derivation.

Since these $J$-factors are galaxy-dependent, once the gamma-ray signals from different galaxies are measured, we can fit the power of $\rho$ and determine the production mechanism of the signal. As is illustrated in \fref{fingerprint}, which assumes that DM follows the Navarro-Frenk-White (NFW) distribution~\cite{Navarro:1995iw}, the two scenarios of canonical annihilation (black) and decay (red) can be distinguished by their ratio of $J$-factors with a reference galaxy\footnote{Throughout this work, we take the Milky Way as our reference galaxy; however, the results can be generalized to other choices of reference.} after taking into account the uncertainty of the NFW fit. We will be using the NFW profile, $\rho(r)=\rho_0\,(r/r_s)^{-1}(1+r/r_s)^{-2}$, for DM halos throughout this paper; however, many of our qualitative results are valid for other choices of the DM profile. In fact, depending on the process of the gamma-ray production from DM, the indirect detection signal can carry a more complex dependence on galactic parameters, such as DM density and halo size, than in \eref{J}. In this work, we discuss the possibility of bringing in such new galactic-dependence in the $J$-factor using the idea of ``non-local" annihilation processes, as explained below. 

As a schematic framework of ``non-local'' annihilation, we consider the possibility of a DM interaction occurring at a given point, $P$, inside the halo first producing a boosted DM particle, see \fref{coordinate}. This boosted particle travels some distance and annihilates with another ambient DM particle producing SM particles at a different location in the galaxy, $P^\prime$, hence dubbed ``non-local.'' As we shall illustrate, due to its mechanism or kinematics, this second annihlation requires the presence of the boosted DM. Not suprisingly, several non-minimal DM models already contain the architecture to include these non-local effects. For instance, such an annihilation process can naturally happen in the semi-annihilation model (see for example~\cite{DEramo:2010keq,Agashe:2014yua}) with asymmetric DM (ADM)~\cite{Kaplan:2009ag,Zurek:2013wia} density in which a boosted DM anti-particle ($\chi^c$) is produced at $P$ from a $\chi\chi$ annihilation via the $\chi\chi\chi X$ coupling (where $X$ is an unspecified particle satisfying $m_X\ll m_{\chi}$). The boosted $\chi^c$ later annihilates with a slow moving $\chi$ at $P'$ giving SM particles through a coupling that contains $\chi\chi^c$. Note that in ADM models, there is no ambient $\chi^c$ for initiating this annihilation, thus requiring production from the first interaction to trigger the second. Of course, the interactions that correspond to each annihilation process are still local.

We define the $J$-factor in the non-local process in a manner analogous to canonical annihilation from \eref{J} with $\sigma_\mathrm{ann}$ and $m_\chi$ substituted for properties of the first annihilation event ($\sigma_1$ and $m_1$). The non-local $J$-factor has additional dependence on the core size and density of the DM halos and the secondary DM annihilation cross-section, the latter being part of the intrinsic particle physics. It is noteworthy that the $J$-factor for the non-local model no longer encapsulates {\em only} astrophysics. This generates another distinct fingerprint in \fref{fingerprint} (blue), with the results depending on an additional product of the galaxy's DM density and size as we discuss below.\footnote{Non-trivial galaxy dependent $J$-factors have been considered in the literature previously, e.g., velocity-dependent DM annihilation~\cite{Boddy:2017vpe,Petac:2018gue,Boddy:2019qak}.}

In order to better illustrate the general concept of non-local annihilation, we first present a toy-model that generates such a non-local annihilation process. The toy-model assumes boosted DM production by another heavier DM annihilation process. It is thus a two component DM model with a two step annihilation process. This is the non-local model shown in \fref{fingerprint}. We then discuss the characteristic features of the non-local $J$-factors in galaxies. Finally, we show an application of the non-local DM annihilation process for explaining the GCE signal and predicting the corresponding gamma-ray signal from the dSph to be smaller than in the canonical model, consistent with observations, unlike the mild tension in the canonical case. Technical details for the $J$-factor calculations are given in the appendices.

\section{DM with non-local annihilation processes}\label{sec.model}

\begin{figure}
\begin{center}
\includegraphics[width=0.95\columnwidth]{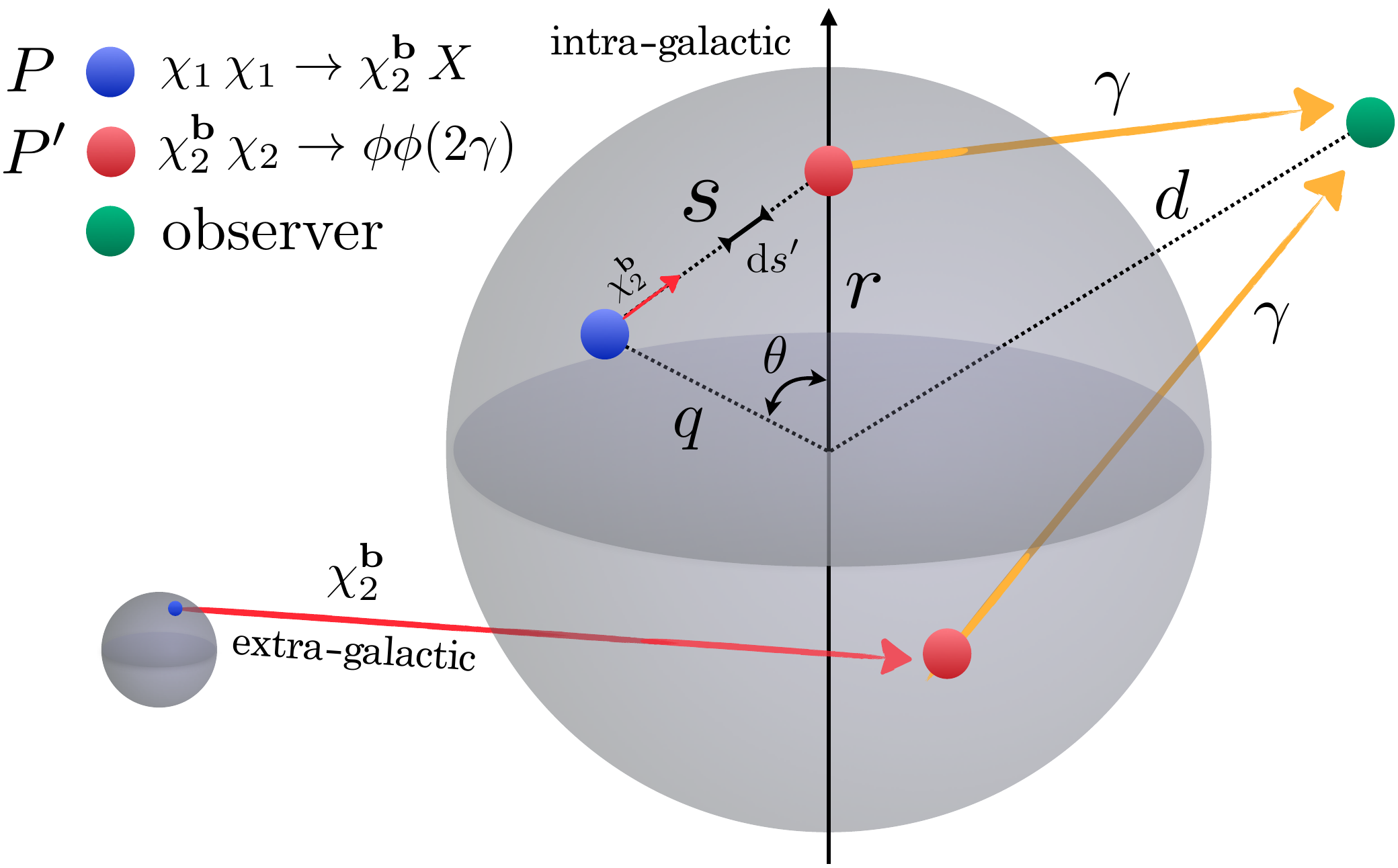}
\caption{An illustration of the non-local annihilation model. A $\chi_1\chi_1$ annihilation first happens at the blue point $P$ a distant $q$ from the halo's center. The produced $\chi_2^{\rm b}$ travels a distance $s$ and annihilates with a slow moving ambient $\chi_2$ at the red point $P^\prime$ into $\phi$'s that decay promptly on galactic scales into gamma-rays which are observed at the green point. $\chi_2^{\rm b}$ can also escape their source galaxy and annihilate in another galaxy as noted by the extra-galactic arrow.}
\label{fig.coordinate}
\end{center}
\end{figure}

We present a concrete toy-model that exhibits the properties of non-local annihilation which were outlined in the introduction. We begin with a summary of the 
general process, followed by a consideration of the motivated parameter space.

In our toy-model, we have a heavy component of DM, denoted by $\chi_1$ annihilating into a lighter DM, $\chi_2$, thus the latter is produced with a boost, being therefore labeled with appropriate superscript, $\chi_2^{\rm b }$:
\begin{equation}
\chi_1\chi_1\to \chi_2^{\rm b}+X
\label{eq.step1}
\end{equation}
The $X$ particle from the first annihilation can either be another dark sector or an SM particle. In this work we simply assume $X$ is an invisible particle that does not participate further in any interactions. This first step is 
followed by $\chi_2^{\rm b}$ annihilating with a stationary $\chi_2$ into another new scalar particle $\phi$:
\begin{equation}
\quad \chi_2^{\rm b} + \chi_2\to 2\phi
\label{eq.step2}
\end{equation}
which ultimately decays into SM particles, namely, photons in our case:
\begin{equation}
\quad\phi\to2\gamma
\label{eq.step3}
\end{equation}
The need for such a mediator between DM and SM will be made clear shortly.

We assume $m_{1}\gg m_{2}$ for the $\chi_{1, \, 2}$ masses, so $\chi_2^{\rm b}$ is relativistic and moves much faster than the escape velocity of the galaxy. We therefore treat all trajectories to be a straight line path, see \fref{coordinate}. In order for the non-local process to be interesting, there is a requirement on the second annihilation cross-section, $\sigma_{2}^{\rm b}$. Once produced, $\chi_2^{\rm b}$ can travel through a typical annihilation length $\ell_{\rm ann}\sim(\sigma_{2}^{\rm b}\,\rho_2/m_2)^{-1}$ before annihilating into $\phi$'s, where $\rho_2$ is the density of the $\chi_2$ background. In order to have a gamma-ray signal to detect, we require a significant fraction, $\gtrsim\mathcal{O}(10\%)$, of $\chi_2^{\rm b}$ to annihilate inside a dSph with radius $\sim{\rm kpc}$. Thus, $\ell_{\rm ann}$ should not be much larger than the halo's characteristic size. We therefore need to satisfy
\begin{equation}
\left(\frac{\rho_2}{10~{\rm GeV\cdot cm}^{-3}}\right)\left(\frac{10~{\rm MeV}}{m_2}\right)\left(\frac{\sigma_{2}^{\rm b}}{(110~{\rm MeV})^{-2}}\right)\gsim 1\,.
\label{eq.large-sigma2}
\end{equation}
Note that in our toy-model presented in \erefs{step1}{step3}, the peak photon energy is approximately $m_1$. Therefore, in order to be within gamma-ray thresholds of the Fermi-LAT experiment, $m_1$ should be in the range $\sim \mathcal{O} (100~{\rm MeV} - 100~{\rm GeV})$. 

In order to produce large boosts, we require $m_2\ll m_1$ and thus by our choice of $m_1\sim \mathcal{O}(100)$ MeV, we are motivated to choose $m_2\sim \mathcal{O}$(10 MeV). The existence of DM particles lighter than $10$~MeV usually encounters strong bounds from the Cosmic Microwave Background (CMB) and Big Bang nucleosynthesis (BBN) measurements (see e.g.,~\cite{Green_2017,Escudero_2019,Depta_2019}). Some studies nevertheless have suggested the possibility of accommodating sub-MeV scale thermal DM with these constraints. For example, Ref.~\cite{Escudero_2019} found that by allowing a small fraction (like $10^{-4}$) of the DM annihilation into neutrinos as compared to $e^+e^-/\gamma$ can alleviate the $\Delta N_{\rm eff}$ and proton-neutron ratio constraints to allow a sub-MeV DM mass. This can help to keep a MeV scale $\chi_2$ without changing the gamma-ray signal significantly. Since our main focus is on the unique feature of gamma-ray signal from the non-local annihilation, we will present results for both $m_2=10$~MeV and $m_2=1$~MeV without specifying the full details of the dark sector that validate the latter case.

The large $\sigma_2^{\rm b}$ cross-section required for the second annihilation has two implications. A direct $\chi_2$ annihilation into photons with such a rate would violate milli-charged DM bounds (see e.g.,~\cite{Berlin_2018,Barkana_2018} and the references therein). We therefore introduce a singlet mediator $\phi$ (see \erefs{step1}{step3}) that has a strong coupling to $\chi_2$ and a suppressed coupling to photons. Secondly, such a large annihilation cross section suggests that $\chi_2$ cannot obtain its relic abundance from a thermal freeze-out process. There are different ways to decouple the $\chi_2$ abundance from its annihilation cross section. For example, in an asymmetric DM scenario, a net $\chi_{1,\,2}$ abundance versus the anti-particles $\chi_{1,\,2}^c$ can be produced from an out-of-equilibrium decay of a heavy particle that strongly violates CP-symmetry. If the heavy particles were produced from a thermal freeze-out process and have an abundance close to the required DM number density, $\chi_{1,\,2}$ can obtain the right relic density\footnote{A similar setup has been discussed in Ref.~\cite{Cui_2013,Cui_2013_2} for baryogenesis.}. After the efficient $\chi_2\chi_2^c$ annihilation depletes $\chi^c$, there is only $\chi_2$ around, and a sizable $\rho_2$ can be obtained inside halos even for a large $\chi_2\chi_2^c$ annihilation.

In order to produce the indirect detection signal in such an asymmetric DM scenario, we consider a more specific model where the two DM particles are complex scalars that carry charges $(-1,+2)$ for $(\chi_1,\,\chi_2)$ under a dark U$_d(1)$ symmetry and have the following couplings:
\begin{equation}
\lambda\,\chi_1\chi_1\chi_2X+C.c.+y_2|\chi_2|^2\phi^2+\hat\lambda |\chi_2|^4+\frac{\phi}{f}F_{\mu\nu}F^{\mu\nu}\,.
\label{eq.Lagrangian}
\end{equation}
In order to simplify the discussion, we only keep couplings that are relevant to the non-local indirect detection signals. The first coupling allows a production of the anti-particle $\chi_1\chi_1\to \chi_2^*+X$ as in \eref{step1}. Here, we consider $\sigma_{1}$ to be similar to the cross section for thermal WIMP DM. Since $\chi_1$ already has the required abundance right after being produced from the out-of-equilibrium decay of a heavy particle (as indicated above, but not explicitly shown in \eref{Lagrangian}), $\chi_1$ annihilation with such a rate is never efficient to significantly change its relic density.

The annihilation cross section of $\chi_2^*\chi_2\to 2\phi$, see \eref{step2}, in the center of mass frame is 
\begin{equation}
\sigma_{2}=\frac{\alpha_2}{4m_1m_2}
\label{eq.crosssemi}
\end{equation}
for $m_\phi\lsim m_2$. We choose $m_1 \sim \mathcal{O}(100)$~MeV and $m_2=\mathcal{O}(1-10)$~MeV, so we need $\alpha_2=y_2^2/4\pi\sim 1$ to obtain a short enough $\ell_{\rm ann}$ for the gamma-ray signal, see \eref{large-sigma2}. Motivated by examples in the lattice studies (e.g.,~\cite{Leino_2018}), we take the perturbativity constraint $\alpha_2\leq1.2$ in this work. Note that a much heavier $\chi_1$ and $\chi_2$ would need larger $\alpha_2$, making the theory non-perturbative.

The large $y_2$ coupling may generate an efficient self-scattering between the ambient $\chi_2$'s through a $\phi$ loop contribution $\frac{\alpha_2}{2\pi}\log\Lambda_{\rm cutoff}$ to the $\hat\lambda|\chi_2|^4$ coupling. In order to satisfy bounds from the various astrophysical constraints, $\sigma_{\chi_2\chi_2\to\chi_2\chi_2}/m_2=\frac{\hat{\lambda}^2}{64\pi m_2^3}\lsim 1$~cm$^2/$g (see~\cite{Tulin_2018} for a review of the bounds), we need the total coupling $\hat\lambda_{eff}\approx\hat{\lambda}+\frac{\alpha_2}{2\pi}\log\Lambda_{\rm cutoff}\lsim(m_2/10~{\rm MeV})^3$. Assuming $\Lambda_{\rm cutoff}\sim 10$ GeV to be larger than all the DM energies we consider, the largest coupling ($\alpha_2 \leq 1.2$) and the lightest $\chi_2$ ($m_2=1$~MeV) require a tuning in $\hat\lambda_{eff}$ no worse than $0.2\%$. After being produced from the $\chi_1$ annihilation, the $\chi_2^{\rm b}$ can also scatter with the ambient $\chi_2$ with cross section $\sigma_{\rm scatt}=\frac{\alpha_{\hat\lambda}}{4m_1m_2}$ and lose its kinetic energy. If the penetration length $\ell_{\rm pen}$ of losing most of the kinetic energy is shorter than $\ell_{\rm ann}$, we cannot assume $\chi_2^{\rm b}$ to fly in a straight line before the annihilation. However, even if $\chi_2^{\rm b}$ loses most of its energy from a single scattering to $\chi_2$ giving $\ell_{\rm pen}\sim (\sigma_{\rm scatt}n_2)^{-1}=\ell_{\rm ann}(\frac{\alpha_{2}}{\alpha_{\hat{\lambda}}})$, $\chi_2^{\rm b}$ annihilation still happens well before the particle slows down for the large $\alpha_2$ we consider.

Finally, $\phi$ couples to photons, see \eref{step3}, via the last term in \eref{Lagrangian}. In principle, $m_{\phi}$ can be larger than $m_2$ as long as the non-local annihilation is kinematically allowed, \eref{step2}. However, in the ADM model that we consider here, we need ambient (non-relativistic) $\chi_2$'s to efficiently annihilate into $\phi's$ to deplete the symmetric part of the $\chi_2$ density, thus we require $m_{\phi}<m_2$ . When showing examples with $m_2\lsim 10$ MeV under this assumption, we need $m_{\phi}<10$ MeV, and the allowed $f$ will be tightly constrained by various bounds on the axion-like-particle, possibly making $\phi$ have a decay length comparable to galactic scales. One way to have $m_\phi \sim \mathcal{O}({\rm MeV})$ while making the $\phi$'s to decay promptly is to consider the cosmological models that can alleviate the $m_{\phi}-f$ bound in the ``cosmological triangle" region, namely $m_{\phi}\sim 1$ MeV and $f\sim10^5$ GeV. For example, as is shown in Ref.~\cite{Depta_2020}, the parameters in the cosmological triangle can be allowed either with the presence of $\Delta N_{\rm eff}$, a non-vanishing neutrino chemical potential, or a lower reheating temperature. In this work, we will present results by assuming $m_{\phi}=1$ MeV and $f=10^5$ GeV without discussing the details of the cosmological model. For larger $m_{1,2,\phi}$, the relevant bounds can easily be satisfied under standard cosmology; however, such mass choices will reduce the gamma-ray signal. For the DM mass we consider, the choice of $\phi$ mass and coupling leads to $\phi$ decay within $10^{-8}$~pc after being produced. The decay is thus prompt compared to galactic sizes.

Note that the non-local behavior of the annihilation still exists even with a smaller coupling and larger $(m_2,m_{\phi})$ that can trivially satisfy the cosmological bounds. Our main motivation for discussing the above scenarios that may require non-standard cosmology is to relate the non-local signal to the known observational sensitivity of the Fermi-LAT experiment. The non-local signal from a simpler dark sector can as well show up in a different energy scale with a different rate.

\section{$J$-factor from the non-local DM annihilation} \label{sec.Jfactor}

Here we study the halo-dependence of the $J$-factor for the non-local (NL) annihilation process, denoted by $J_{\rm NL}$. There are two main sources of $\chi_2^{\rm b}$ involved in the secondary annihilation. $\chi_2^{\rm b}$ can either come from a $\chi_1$ annihilation inside the same halo (``intra-galactic'', IG) or from a $\chi_1$ annihilation in another galaxies (``extra-galactic'', EG). The two types of signal carry different dependence in DM density. We therefore have 
\begin{equation}
J_{\rm NL}\approx J_{\rm IG}+J_{\rm EG}
\end{equation}
for the non-local $J$-factor. There are also signals coming from $\chi_2^{\rm b}$ produced in the inter-galactic region, but the signal rate is negligible due to the low DM density outside of galaxies.

In the limit of large $\sigma_2^{\rm b}$, $J_{\rm NL}$ is dominated by the intra-galactic contribution and reproduces the galactic-dependence from the canonical DM annihilation scenario. Whereas, in the other extreme of small annihilation cross-section, both the intra- and extra-galactic sources contribute. The intra-galactic contribution, $J_{\rm IG}$, behaves similar to the canonical DM annihilation with an additional galaxy dependent modulation factor. The extra-galactic contribution, $J_{\rm EG}$, has the galactic-dependence of the decay DM scenario. Dominance of intra- versus extra- depends on galactic parameters with larger galaxies favoring the intra-galactic contribution and vice versa. In this section, we will demonstrate these expected results explicitly. To avoid confusion, we will refer to the $J$-factors for canonical annihilation and decay, as given in \eref{J}, by $J_{\rm ann}$ and $J_{\rm dec}$, respectively.

\subsection{Annihilation from intra-galactic $\chi_2^{\rm b}$}

We first present the expression for $J_{\rm IG}$, then provide some intuition behind it based on simplifying assumptions. Details of the derivation are given in \aref{J_NL_derivation}. After first defining the coordinates as in \fref{coordinate}, $J_{\rm IG}$ can be written as
\begin{eqnarray}
J_{\rm IG} & = & \int_{\rm ROI}\, {\rm d}\Omega_\ell \int_{\rm los} {\rm d} \ell \label{eq.J_nl_exact} \\
& \times & \int_{\vec{s}}\frac{{\rm d}^3\vec{\hat{s}}}{\hat{s}^2}\frac{{\rm d}{\mathcal P}_{\chi_2^{\rm b}\chi_2}(\hat{r},\hat{s})}{{\rm d}\hat{s}}\,[\rho_{1,0}\eta_1(\hat{q})]^2 \frac{{\rm d}N}{{\rm d}\Omega_{\vec{s}}}(\vec{\hat{s}},\vec{\ell})\,,\nonumber
\end{eqnarray}
where the integration is performed over a region-of-interest (ROI) and a line-of-sight (los). In order to better identify the galaxy-dependent parameters in the expression, we define the dimensionless lengths $(\hat{r},\hat{s},\hat{q})=(r,s,q)\,r_s^{-1}$ so that the integral over the lengths is independent of the galaxy's size. In this notation, the $\hat{q}$ is a function of $(\hat{r},\theta,\hat{s})$ as in \fref{coordinate}. We also define $\eta_i(\hat{r})=\rho_i(r_s \, \hat{r})/\rho_{i,\,0}$ to separate the galaxy-dependent properties from the characteristic profile, where $i$ corresponds to $\chi_{i=1,2}$. Here, ${\mathcal P}_{\chi_2^{\rm b}\chi_2}$ is the probability of having $\chi_2^{\rm b}$ annihilate after traveling a displacement $\vec{s}$ from the first ($\chi_1$) annihilation point
\begin{equation}\label{eq.prob}
\begin{aligned}
\frac{{\rm d}{\mathcal P}_{\chi_2^{\rm b}\chi_2}(\hat{r},\hat{s})}{{\rm d}\hat{s}}=\Lambda\,\eta_2(\hat{r})
\exp\left[-\Lambda \int_0^{\hat{s}}{\rm d}\hat{s}'\,\eta_2(\hat{s}')\right]\,,
\end{aligned}
\end{equation}
where $\int{\rm d}\hat{s}'$ integrates the annihilation probability $\chi_2^{\rm b}$ on its way to the final annihilation point. The probability function is solely dependent on the characteristic density profile and the  dimensionless quantity
\begin{equation}
\label{eq.Lambda}
\Lambda\equiv r_s\,\rho_{2,0}\,\sigma_{2}^{\rm b}/m_2
\end{equation}
which is roughly just the inverse of the typical annihilation length ($\ell_{ \rm ann }$) introduced in the earlier section in units of the halo/core size. In the case of $r_s \ll \ell_{ \rm ann }$, it corresponds to the probability of $\chi_2^{\rm b}$ annihilating inside a halo with constant $\chi_2$ density $\rho_{2,0}$ and characteristic size $r_s$. The exponential factor in \eref{prob} indicates the surviving probability of $\chi_2^{\rm b}$ after traveling a distance $s$ to the second annihilation point. ${\rm d}N/{\rm d}\Omega_{\vec{s}}$ is the angular distribution of the signal as a result of the second annihilation occurring in a boosted frame and is dependent on the angle between the direction of  $\chi_2^{\rm b}$'s momentum, $\vec{r}$, and the observer, $\vec{\ell}$. In order to write it in the form shown in \eref{ndep1}, we assumed the spectrum does not depend on this angle. This is supported by our assumption discussed later of approximating the angular distribution with a delta function. For a more complete equation including spectral angular dependence, see \aref{J_NL_derivation}. However, in the limit $d\gg r_s$ where $d$  is the distance the galaxy is away from the observer, this effect can be approximated as effectively isotropic. This isotropy is a result of all points in the galaxy being equally far from the observer, resulting in the various $\chi_2^{\rm b}$ directions averaging out over the final volume integral. In this far away galaxy approximation,
\begin{equation}\label{eq.realJ}
J_{\rm IG}=d^{-2}\int_{\vec r}{\rm d}V\int_{\vec{s}}\frac{{\rm d}^3\vec{\hat{s}}}{4\pi\hat{s}^2}\frac{{\rm d}{\mathcal P}_{\chi_2^{\rm b}\chi_2}(\hat{r},\hat{s})}{{\rm d}\hat{s}}\,[\rho_{1,0}\eta_1(\hat{q})]^2.
\end{equation}
For detailed calculations of $J$-factors in this work, we use the full expression \eref{J_nl_exact} assuming ${\rm d}N/{\rm d}\Omega_{\vec{s}}$ is a delta function in line with $\vec{s}$ due to the high boost of $\chi_2^{\rm b}$ in the second annihilation.

Next, we consider two limiting cases of $J_{\rm NL}$ through $\Lambda$ in order to understand analytically the morphology of the NL signal versus the canonical annihilation scenario. Recall from \eref{Lambda} that $\Lambda$ is effectively the inverse of the free-streaming length of $\chi_2^{\rm b}$ in units of galactic size. This also serves as a useful cross-check.

In the $\Lambda\gg 1$ limit, it is clear that 
$\chi_2^{\rm b}$ annihilates right after its production from the $\chi_1$ annihilation. The exponential factor in \eref{prob} is non-negligible only for $s \lesssim r_s / \Lambda \ll r_s$. We thus expect the $J$-factor for the NL model to be proportional to $\rho_1^2$ as in \eref{J} for the canonical case. Indeed, by taking the large $\Lambda$ limit in \eref{prob}, since $\displaystyle{\lim_{\Lambda\eta_2\to\infty}}\Lambda\eta_2\exp(-\Lambda\eta_2\hat{s})\approx\delta(\hat{s})$ for $\hat{s}\geq 0$, \eref{realJ} recovers the result for the canonical annihilation process:
\begin{equation}
J_{\rm IG} = d^{-2}\int {\rm d}V \rho_1^2(r)
\label{eq.J_IG_short_free_stream}
\end{equation}

On the other hand, when $\Lambda\ll 1$, the exponential factor in \eref{prob} reduces to one if we expand the expression to linear order in $\Lambda$. It is also convenient to perform the volume integral $\int{\rm d}^3\vec{s}$ in terms of $\int{\rm d}^3\vec{q} \sim 4 \pi \int {\rm d} q \, q^2$. \eref{realJ} thus reduces to
\begin{equation}\label{eq.smallL_exact}
J_{\rm IG}=4\pi \, \rho_{1, \, 0}^2 \, r_s^3 \, d^{-2} \Lambda \int\hat{r}^2{\rm d}\hat{r}\, \eta_2(\hat{r})\,\int\frac{\hat{q}^2{\rm d}\hat{q}}{\hat{s}^2}\,[\eta_1(\hat{q})]^2.
\end{equation}
where the integrals are dimensionless and only depend on the characteristic profile. They are thus identical for all galaxies with the same profiles $\eta_i$. Following the assumption of the NFW profile, we can further relate this expression to $J_{\rm ann}$ in the canonical annihilation case. Since the gamma-ray signal is mainly produced in the inner part of the halo (so $\hat r,\hat q\lsim 1$), the integral gets its dominant contribution when the DM profile is $\eta_1(x)=\eta_2(x)\sim x^{-1}$ for $x=\hat r$ or $\hat q$. The $\hat{s}$ in the integrand is approximately $\hat s\sim \hat r$ when $0\lsim\hat q\lsim\hat r$, and $\hat s\sim \hat q$ when $\hat r\lsim \hat q\lsim 1$. After performing the ${\rm d}\hat{q}$ integral for $0\lsim\hat q\lsim 1$ and using the relation between DM profiles, we can rewrite the $J$-factor as
\begin{eqnarray}\label{eq.smallL}
J_{\rm IG}&\sim& \Lambda\,d^{-2}\int {\rm d}V\,\rho_{1}^2(r)\sim\Lambda J_{\rm ann}.
\end{eqnarray}
The result is rather intuitive since it is the $J_{\rm ann}$ in Eq.~(\ref{eq.J}) that initiates the process from a canonical $\chi_1$ annihilation times a suppression factor $\Lambda$ that corresponds to the probability of $\chi_2^{\rm b}$ annihilation. Under the same assumption of the NFW profile and the isotropy of the signal, a similar estimate can be done for the MW, and it can be shown that the $J$-factor is also $\sim\Lambda J_{\rm ann}$ but with $\Lambda$ derived for the MW halo. Thus, NL annihilation produces an additional $\rho_{2,0} \, r_s$ dependence via $\Lambda$ to the $J$-factor that is not present in the canonical framework. This additional term is a galaxy-dependent modulation to the $J$-factor. In \fref{fingerprint}, we have therefore ordered the dSphs in the horizontal axis by increasing $\rho_0 \, r_s$, see \tref{galaxy_param}. As we can see, the variations of the $J$-factor ratios from canonical annihilation do indeed follow the same ordering.

\begin{table}
\center{\begin{tabular}{|c|c|c|c|}
\hline
\,Galaxy\, & \,$\rho_0$~$\left[{\rm GeV/cm}^{-3}\right]$ \, & \,$r_s$~[kpc] \,&\,\, $\frac{\left(\rho_0\,r_s\right)^{\rm MW}}{\left(\rho_0\,r_s\right)^{\rm Gal.}}$\,
\\[3pt] \hline
MW & $0.345$ & $20$ & $1$ \\ \hline
Sextans & $0.218$ & $2.10$ & $15.1$ \\ \hline
\,Canes Venatici I\, & $0.381$ & $1.70$ & $10.7$ \\ \hline
Fornax & $0.359$ & $2.44$ & $7.89$ \\ \hline
Carina & $1.18$ & $0.812$ & $7.22$ \\ \hline
Leo I & $1.13$ & $1.17$ & $5.25$ \\ \hline
Sculptor & $1.74$ & $0.920$ & $4.33$ \\ \hline
Leo II & $2.57$ & $0.636$ & $4.23$ \\ \hline
Ursa Minor & $2.54$ & $0.804$ & $3.38$ \\ \hline
Draco & $2.96$ & $0.728$ & $3.20$ \\ \hline
\end{tabular}}
\caption{Best fit galactic halo density and radius parameters for various galaxies. The ratios of $\rho_0\,r_s$ for each galaxy with the Milky Way are also shown. The DM distribution is assumed to be the NFW profile. Values for the Milky Way are derived using a local density of 0.4 GeV/cm$^{-3}$, $r_s=20$ kpc, and our local radius of 8.5 kpc. The dSph values are derived from Ref.~\cite{Pace:2018tin}.}\label{t.galaxy_param}
\end{table}

Traditionally, $J$-factors are independent of particle physics such as mass and cross-section. In order to keep a consistent definition of $J_{\rm IG}$ for all scales in \eref{J_nl_exact}, we have left the $\sigma_2^{\rm b}$ dependence in \erefs{smallL_exact}{smallL}, but the cross-section is separable. However, except in the most extreme cases of \eref{J_nl_exact} as observed in \eref{J_IG_short_free_stream} and \eref{smallL_exact}, the secondary annihilation cross-section is genuinely inseparable from the astrophysics. This region corresponds to the critical value of $\Lambda \sim 1$ where we transition between these two extreme cases.

\begin{table}
\center{\begin{tabular}{|c|c|c|c|}
\hline
\,Model\, & \,$m_1$~[GeV] \,& \,$m_2$~[MeV] \,&\,\, $f_2$\,\, \\ [1pt] \hline
A & $0.1$ & $10$ & $0.9$ \\ \hline
B & $0.1$ & $1$ & $0.1$ \\ \hline
C & $0.1$ & $1$ & $0.9$ \\ \hline
GCE & $5.68$ & $1$ & $0.1$ \\ \hline
\end{tabular}}
\caption{DM masses and energy density fraction used in the different example models. The $m_1$ of the ``GCE" case comes from fitting the gamma-ray spectrum to the GCE signal. As discussed in Sec.~\ref{sec.model}, we choose $m_{\phi}=1$ MeV in all the examples. We assume the NFW profile $\rho_1/\rho_{1,0}=\rho_2/\rho_{2,0}=[r/r_s(1+r/r_s)]^{-1}$ for all the models.}\label{t.param}
\end{table}

In this work, we present numerical results for the four example models described in \tref{param}. Besides the different choices of DM masses, we also keep the fraction of $\chi_2$  density as a variable
\begin{equation}
f_i\equiv\frac{\rho_{0,i}}{\rho_{0,1}+\rho_{0,2}}\,,\quad i=1,2
\end{equation}
and assume both particles follow the same NFW profile in all cases. One can relax the assumption and follow the same analysis as we describe for different DM profiles. In \fref{GCE0} (left), we show an example of the constraints using Model A that can be placed on the NL process described in \erefs{step1}{step3} in the $\sigma_1-\sigma_2^{\rm b}$ plane by requiring the gamma-ray flux to be constant.\footnote{We require the flux be described by \eref{PhiGCE} with $f_{0} = 9.38\times 10^{-8}\, {\rm cm}^{-2}\, {\rm s}^{-1}\, {\rm sr}^{-1}$ described later in this work. This particular value for $f_0$ is the best fit spectral normalization for our toy-model to the GCE which requires $m_1=5.68$~GeV.} We rescale the required annihilation cross-sections shown in the axes labels by $f_{1,2}$ and $m_{1,2}$, so the result (black) curve is the same for all the \tref{param} models. This is observed in \fref{GCE0} (right) where the only difference between the various scenarios is the CMB and the perturbativity constraints. The CMB bound (dashed black line) on the photon injection from $\chi_2$ annihilation assumes the second annihilation is prompt around reionization due to increases in $\chi_2$'s density. The CMB bound requires $\langle \sigma_{1} v\rangle \, \lsim \, 2 \times 6\times 10^{-26}\,(m_1/7~{\rm GeV})\,(1-f_2)^{-2} \,{\rm cm}^3/$s~\cite{Slatyer:2015jla,Aghanim:2018eyx}. We therefore set a lower bound on $\sigma_{2}^{\rm b}/m_2$ by requiring that the lower 1$\sigma$ error bar on $\sigma_1$ needed to fit the flux be below the CMB bound. The factor of 2 and $f_2$ are a result of rescaling to account for only half of the annihilation energy going into SM particles and a different $\rho_1$, respectively. For the DM masses we consider, the $\alpha_2$ coupling in \eref{crosssemi} becomes non-perturbative when $\sigma_{2}^{\rm b}/m_2>3\times 10^4$~GeV$^{-3}$; this sets an upper bound on the $\chi_2^{\rm b}$ annihilation. The allowed range of $\sigma_{2}^{\rm b}/m_2$ is displayed in purple.

\begin{figure*}
\includegraphics[width=1\columnwidth]{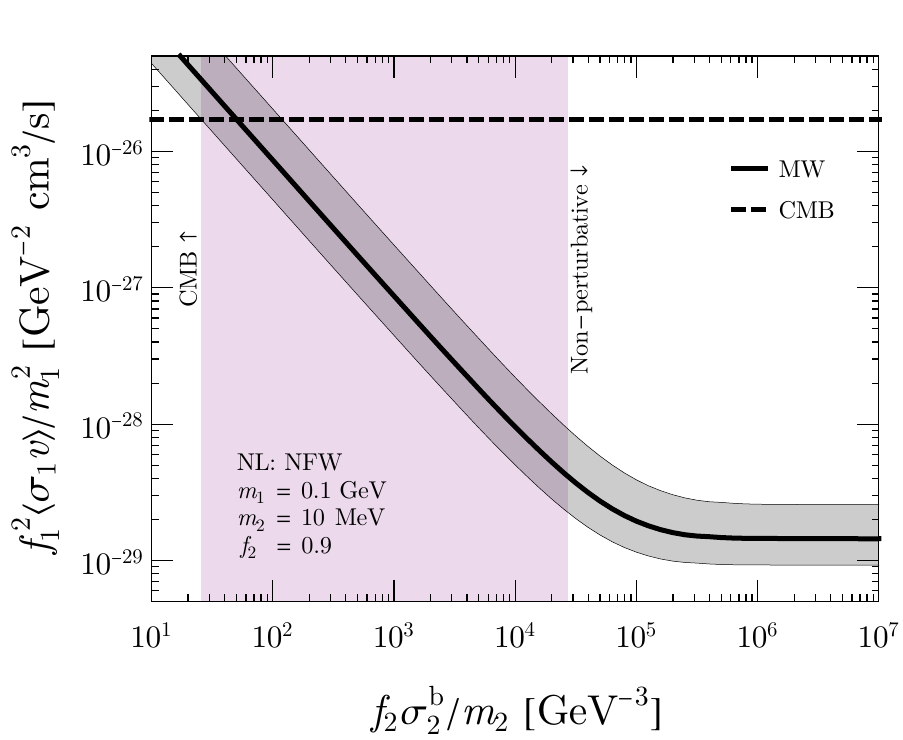}\,\,\,\,\,\,\includegraphics[width=1\columnwidth]{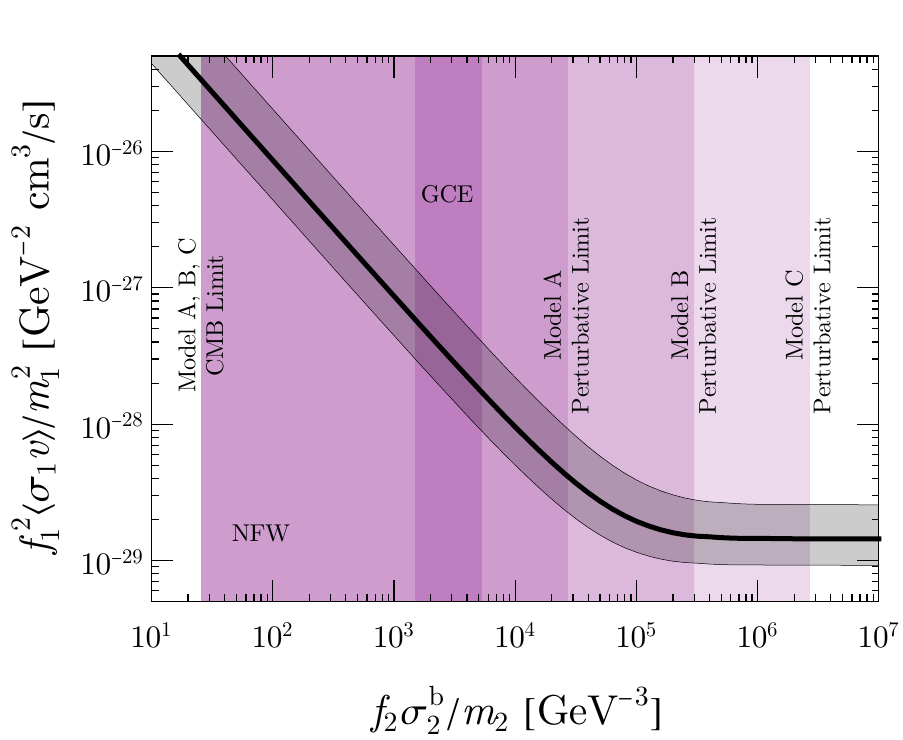}
\caption{The $\chi_1$ annihilation rate in the NL model (solid) for producing gamma-ray signals consistent with a fixed flux in the MW for Model A (left) and a combined image of all example models (right). The width of the band corresponds to 1$\sigma$ error bars assuming the local $\rho_{\rm MW} = 0.4 \pm 0.1$~GeV/cm$^3$ and $r_{\rm Earth}=8.5$ kpc. The required $\langle\sigma_1 v\rangle$ to fit the flux decreases linearly for $\sigma_2^{\rm b}/m_2$ that is much smaller than the critical value of $\sigma_2^{\rm b}/m_2$ corresponding to $\Lambda\sim1$, see text for details. At larger $\sigma_2^{\rm b}/m_2$ where $\Lambda \gg 1$, $\langle\sigma_1 v\rangle$ is constant as the MW exits the non-local regime. The CMB upper bound (dashed) on $\langle\sigma_1 v\rangle$ for the model is also shown, and we translate it into a lower bound on $\sigma_2^{\rm b}$ for a fixed flux (the intersection between the solid and dashed curves). Model masses are shown in \tref{param} as benchmark examples. We take $\alpha_2\leq1.2$ as the non-perturbativity constraint and set an upper bound on $\sigma_2^{\rm b}$ via \eref{crosssemi} . Allowed regions for $\sigma_2^{\rm b}/m_2$ are shaded in purple with the left edge set by violating CMB constraints and the right by the model becoming non-pertubative.}
\label{fig.GCE0}
\end{figure*}

The signal in \fref{GCE0} originates from the MW with an ROI $2^\circ<\theta<20^\circ$ from the galactic center, and we only consider the intra-galactic contribution; however, the extra-galactic contribution is negligible for the MW as shown later. Note that, even though we use a normalization influenced by the GCE to obtain these results, the choice of $m_1$ for Models A, B, and C produces a $E_\gamma^2{\rm d}N_{\gamma}/{\rm d}E$ spectrum peaked around $50$~MeV and thus cannot explain the GCE; however, the resulting gamma-rays are still energetic enough to be potentially observable in the future.

The requirement of the signal flux determines $\langle \sigma_{1} v\rangle$ in \fref{GCE0} as a function of $f_2\sigma_{2}^{\rm b}/m_2$. The relevant $J$-factors are calculated by numerically solving \eref{J_nl_exact} for the galactic signal. As anticipated from \eref{smallL}, the MW signal for the NL model is linearly suppressed for small $\Lambda=(\rho_{2,0}\,r_s)\,\sigma_{2}^{\rm b}/m_2$, thus necessitating a larger $\langle \sigma_{1} v\rangle$ to obtain the required signal rate. The NL suppression no longer applies for $\Lambda \gsim 1$, i.e., when $f_2\,\sigma_{2}^{\rm b}/m_2 \gsim 10^5$~GeV$^{-3}$ for the MW. Thus, the $J$-factor asymptotes to the canonical DM annihilation as we have discussed, and the required $\langle \sigma_{1} v\rangle$ no longer depends on $\sigma_{2}^{\rm b}/m_2$.

Next, using our toy NL model, we study the signal rate for different dSphs as a function of $\sigma_{2}^{\rm b}/m_2$. In \fref{GCE3}, we show the ratio of $J$-factors between dSph's and the MW signals using the same ROI as \fref{fingerprint}. The solid and dashed curves are from the ``non-local" and the ``canonical" DM annihilations for Sextans (blue) and Draco (red). In the small $\chi_2^{\rm b}/m_2$ annihilation limit, we observe a clear reduction of the ratio between the non-local signals compared with their canonical counterpart due to both MW and dSph annihilations suffering a similar $\Lambda\ll1$ suppression. From the discussion below Eq.~(\ref{eq.smallL}), the ratio of the $J$-factors for dSph vs.~MW is modified in the non-local model relative to the canonical by $\sim\Lambda^{\rm dSph}/\Lambda^{\rm MW}=\left( \rho_{2, \, 0} \, r_s \right)^{\rm dSph} / \left( \rho_{2, \, 0} \, r_s \right)^{\rm MW}$. Crucially, $\rho_{2, \, 0} \, r_s$ is different for each galaxy with the MW being the largest in our local group by a factor of a few which explains the suppression of $J$-factor ratios shown in \fref{GCE3}, see \tref{galaxy_param}. Moreover, this effect varies with the specific dSph in consideration; it is however independent of both $\chi_{1}$ and $\chi_{2}$ particle parameters. Indeed, as seen in \fref{GCE3}, this dilution is more significant for Sextans because it's $\rho_{2,0}\, r_s$ is smaller than Draco's. This small cross-section regime is what is plotted in \fref{fingerprint}. The $J$-factor suppression is clearly seen and its magnitude decreases as we move horizontally on the figure to larger $\rho_{2,0}\,r_s$, matching the above expectation.

\begin{figure*}
\includegraphics[width=1.3\columnwidth]{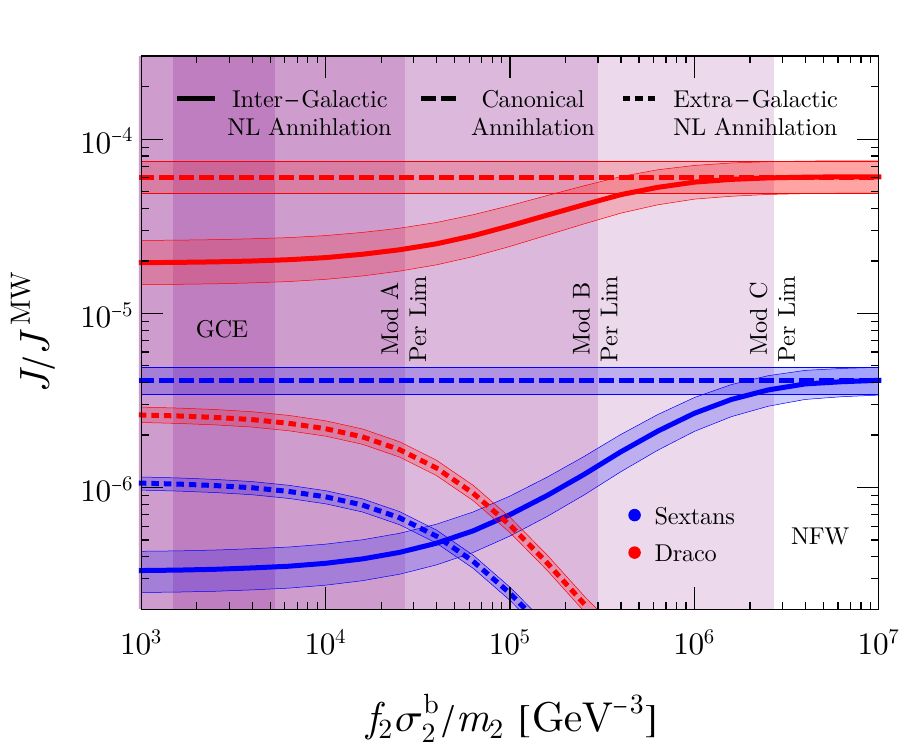}
\caption{Ratio of $J/J^{\rm MW}$ for select dSphs. The non-local $J$-factor (solid) is constant at large $\sigma_2/m_2$, drops when the dSph enters the non-local regime, then levels out when MW also becomes non-local. The canonical annihilation case is also shown (dashed). These two models are the same at large $\sigma_2/m_2$ as they are both local. Estimates for the extra-galactic non-local contribution from each galaxy are also shown (dotted). The bands are 1$\sigma$ error estimates. The error bars for the extra-galactic portion only reflects errors in the second annihlation galaxy, and does not include any portion from uncertainties in determining the background $\chi_2^{\rm b}$ flux from the extra-galactic sources. We also show the allowed regions due to CMB and non-perturbativity bounds discussed in \fref{GCE0} using purple shade for the four models in \tref{param}.}
\label{fig.GCE3}
\end{figure*}

As already indicated in \fref{GCE0} for MW, upon increasing $\sigma_2^{ \rm b}$, the galaxies start exiting from the NL suppression and become canonical for $\sigma_2^{ \rm b }/m_2 \gsim 1 / \left( \rho_{2, \,0} \, r_s \right)$. At this scale, the annihilation length of $\chi_2^{\rm b}$ is smaller than the galaxy, and the $J$-factor eventually asymptotes to the canonical result. This transition from NL to canonical gives rise to interesting features in the ratio of $J$-factors. Because each galaxy has a different $\rho_{2,0} \, r_s$, they each transition at a different $\sigma_2^{ \rm b }/m_2$. This behavior is directly observed in \fref{GCE3}. The rise in the ratios of $J$-factors around $10^5$~GeV$^{-3}$ corresponds with MW's transition while the flattening of the ratio around $10^6$~GeV$^{-3}$ corresponds to each dSph's transition. Again note that the particular ordering and scale of the flattening of the $J$-factor ratio for the two galaxies corresponds to the hierarchy in $\rho_{2,0}\,r_s$.

Conceptually, it is convenient to consider these two transitions and the three distinct regions they produce with decreasing cross-section moving from right-to-left on \fref{GCE3} in contrast to our earlier discussion which moved from left-to-right. For large $\sigma_2^{ \rm b }/m_2$, we identify the canonical region where both the MW and the dSph are in the $\Lambda\gg1$ regime; here, their $J$-factors do not depend on the second particles properties and are thus constants. Note that the ratio merges with the canonical annihilation ratio as expected. As we lower the cross-section, because $\rho_{2,0} \, r_s$ of Sextans is smaller than Draco, Sextans exits the canonical regime at a slightly larger $\sigma_2^{ \rm b }/m_2$ than Draco, as seen in \fref{GCE3}. Next, the intermediate region where the galaxy with smaller $\rho_{2, \, 0}\,r_s$ becomes NL while the other is still canonical. This results in the $J$-factor ratio having linear dependence on $\sigma_2^{ \rm b }/m_2$ via $\Lambda^{\rm dSph}$. Finally, the pure NL region where both galaxies are NL, each galaxy has its own $\Lambda$ dependence. This results in the ratio $\sim \left( \rho_{2, \, 0} \, r_s \right)^{\rm dSph} / \left( \rho_{2, \, 0} \, r_s \right)^{\rm MW}$, independent of $\sigma_2^{\rm b}$, as discussed earlier.

In summary, the intra-galactic NL contribution possesses a striking feature as seen in \fref{fingerprint} and \fref{GCE3}. For a fixed MW flux, not only is there a suppression of the signal relative to the canonical model for each dSph, but the level of suppression depends on the density and size of the galaxy as well as the $\chi_2^{\rm b}$ annihilation cross-section, as shown in \fref{GCE3}.

\subsection{Annihilation from extra-galactic $\chi_2^{\rm b}$}

Since most of the $\chi_2^{\rm b}$s can escape their source galaxy in the $\Lambda\ll 1$ limit, we should also consider $\chi_2^{\rm b}$ produced in {\em other} galaxies traveling to and annihilating in a given target galaxy. As we will discuss, the signal produced by extra-galactic $\chi_2^{\rm b}$ has a $J$-factor halo-dependence similar to the decay DM scenario unlike the intra-galactic discussed above. The extra-galactic signal magnitude is roughly comparable to the intra-galactic signal, either can be the dominant contributor depending on the number of halos in the universe which produce the extra-galactic $\chi_2^{\rm b}$ flux and the characteristics of the target galaxy. Larger galaxies are more likely to be intra-galactic dominated due to their large internal $\chi_2^{\rm b}$ production.

We first provide an order of magnitude estimate of the signal rate. We assume the $\chi_2^{\rm b}$ flux to be mainly produced from MW-sized main galaxies (MG) that are uniformly distributed throughout the whole universe. We take the average galactic mass to be $\sim 8\times 10^{11}M_{\odot}$ based on the Virgo Cluster.\footnote{Estimates on the Virgo cluster assume a mass $M_{\rm Virgo}=1.2\times 10^{15}M_{\odot}$~\cite{Fouque:2001qc} and a galaxy count $N_{\rm Galaxies, \, Virgo} = 1500$~\cite{Binggeli:1985zz,Binggeli:1987qv}.} This mass is near MW's supporting our $\chi_2^{\rm b}$ estimate. With the average matter density in the universe\footnote{The average matter density is based on $h=0.7$ and $\Omega_m=0.3$.} $\rho_m= 4.1\times10^{10}M_{\odot}/$Mpc$^{3}$, we estimate the average galaxy density $n_{\rm halo}\sim 0.05$ Mpc$^{-3}$.

We assume $\Lambda \ll 1$, and most $\chi_2^{\rm b}$ leave their source galaxy, so the rate of $\chi_2$ annihilation in a nearby ``target" galaxy (dubbed TG) that we observe is given by
\bea
\label{eq.signalEG}
\frac{{\rm d}N_{\chi_{2}}^{\rm ann}}{{\rm d}t} & \sim &
\frac{\langle\sigma_1 v\rangle}{2\,m_1^2} \, \Phi_{\rm halo} \left[\int {\rm d}\bar{V}\,\frac{n_{\rm halo}}{4\pi \bar{r}^2} \right] \left( \pi r_s^2 \Lambda \right)^{\rm TG}\quad
\eea
The leading terms in front of the square brackets in general estimate the production rate of $\chi_2^{\rm b}$ and their escape probability from a single main galaxy. Since here we are working in the $\Lambda\ll 1$ limit where most $\chi_2^{\rm b}$'s escape their parent galaxy, it reduces to simply the $\chi_2^{\rm b}$ production rate with
\begin{equation}
\Phi_{\rm halo} \approx \int {\rm d}V (\rho_1^{\rm MG}(r))^2.
\label{eq.Phi_halo}
\end{equation}
The number density integral in square-brackets estimates the total number of halos in the visible universe with an area suppression which accounts for dilution of $\chi_2^{\rm b}$ flux due to distance from the target galaxy. The final term is the capture cross-section of the target galaxy, $(\pi r_s^2 \Lambda)^{\rm TG}$, being the physical area multiplied by the probability of capture. The corresponding $J$-factor can thus be obtained through an ROI and los integration: $J_{\rm ann} = \frac{2m_\chi}{\langle\sigma v\rangle}_\chi \, \int_{\rm ROI} \int_{\rm los} {\rm d}\ell \, {\rm d} \Omega \, \frac{{\rm d}N_{\chi}^{\rm ann}}{{\rm d}t}$. In the $d \gg r_s$ and $\Lambda \ll 1$ limits,
\begin{eqnarray}
J_{\rm EG} & \sim & \left( \pi n_{\rm halo}\,R \left(r_s^2\right)^{\rm TG} \right) \frac{ \Lambda^{\rm TG} }{ d^2 } \int {\rm d}V \left[\rho_1^{\rm MG}(r)\right]^2 \nonumber \\ 
& \sim & \left( \pi n_{\rm halo}\,R \left(r_s^2\right)^{\rm TG} \right) \frac{\left(\rho_{1,0}^2 \, r_s^3\right)^{\rm MG}}{\left(\rho_{1,0}^2 \, r_s^3\right)^{\rm TG}} \left( \Lambda J_{\rm ann}\right)^{\rm TG}\quad\;
\label{eq.JEG}
\end{eqnarray}
where $R$ is the radius of the visible universe from which $\chi_2^{\rm b}$ originate.\footnote{Note that for $J_{\rm EG}$ calculations, the extra-galactic volume integration is performed in comoving coordinates. Our constant $n_{\rm halo}$ therefore naturally includes factors related to expansion of the universe when working in other coordinates. However, we do not include any additional alterations to the halo population. We estimate inclusion of changes to the halo population to be less than an order of magnitude correction to our result due to the growth of virial overdensity as a function of redshift~\cite{Klypin:2010qw,Allahverdi:2011sx}. Additionally, an interesting outcome of the expansion which we omitted in this calculation is the redshift dependence of $\chi_2^{\rm b}$'s energy which would result in an altered gamma-ray spectra. We leave a more detailed analysis of these redshift dependent effects to an upcoming work \cite{NL_local_in_progress}.} We use the usual expression for canonical annihilation $J_{\rm ann}$ from \eref{J} and simply re-write the result such that the final parenthesis is similar to the $J$-factor estimate from the intra-galactic contribution in \eref{smallL} for ease of comparison.

The $J_{\rm EG}$ carries an additional suppression relative to the intra-galactic contribution of $\pi n_{\rm halo} R \, (r_s^2)^{\rm TG}\sim 10^{-3}$, where $(r_s)^{\rm TG}\sim$ kpc is the typical size of dSphs and we take $R=9$ Gpc for the distance back to redshift $z\approx 8$ at the reionization and assume the opacity factor to be $1$~\cite{Allahverdi:2011sx}. On the other hand, the middle term in Eq.~(\ref{eq.JEG}) gives an enhancement for a target galaxy smaller than the typical main galaxies. This term originates from the conversion of the galactic volumetric integral which characterizes the $\chi_2^{\rm b}$ production rate from a main galaxy. For dSph, this ratio is of $\mathcal{O}(10^{3})$.

Combining all these factors, we see that the resulting $J_{\rm EG}$ is of the same order of magnitude as the intra-galactic \eref{smallL} for dSph's and sub-dominant for MG sized galaxies. For extra-galactic NL annihilation, since we integrate over the $\chi_2^{\rm b}$ source galaxies in the whole universe, the only halo dependence originates from the target galaxy yielding $J_{\rm EG}\propto \rho_{2, \, 0}/m_2\,r_s^3/d^2$, which has similar galactic dependence as the $J$-factor for decaying dark matter, see \eref{J}.

A more exact calculation of the extra-galactic contribution can be derived from \eref{J_nl_exact} by substituting $(\rho_{1,0}\eta_1)^2\to n_{\rm halo}\,\Phi_{\rm halo}$. This substitution alters the production method for $\chi_2^{\rm b}$. Instead of being produced inside the target galaxy, they are now produced uniformly from all space. This simulates an average background of $\chi_2^{\rm b}$s that are produced inside and escape from main galaxies throughout the universe.

In order to account for a possible $\chi_2^{ \rm b}$ annihilation in the ``inter-galactic" medium, $\Lambda\eta_2$ in the exponential of \eref{prob} is replaced with $\Lambda^{\rm InterG}$ where InterG denotes the inter-galactic values (note that $\eta_2^{\rm InterG}=1$) making the integration trivial.\footnote{Note that even though $r_s^{\rm InterG}$ from $\Lambda^{\rm InterG}$ has no physical connection to the target galaxy, it should remain as $r_s^{\rm TG}$ in order to maintain a consistent definition for the dimensionless integration variables, see the discussion below \eref{J_nl_exact}.} With these changes, extra-galactic $J$-factor becomes
\begin{equation}\label{eq.JEGfull}
J_{\rm EG} \approx R_{\rm eff} \, n_{\rm halo} \, \Phi_{\rm halo}\left(\frac{\Lambda
}{r_s}\right)^{\rm TG} \int_{\rm los} \int_{\rm ROI} {\rm d}\ell \, {\rm d}\Omega \, \eta_2^{\rm TG}(\hat{r})
\end{equation}
with
\begin{equation}
\label{eq.Reff}
R_{\rm eff} = \ell_{\rm ann}^{\rm InterG} \left( 1 - e^{-R/\ell_{\rm ann}^{\rm InterG}}\right) 
\end{equation}
where $\ell_{\rm ann}^{\rm InterG} = m_2/(\sigma_2^{\rm b} \, \rho_2^{\rm InterG} )$ is the typical annihilation 
length in the inter-galactic medium and $R$ is the same from \eref{JEG}. $R_{\rm eff}$ originates from $\chi_2^{\rm b}$ suppression due to inter-galactic annihilations integrated over the entire volume. In the limit that omits inter-galactic annihilations $R_{\rm eff}$ becomes $R$. Furthermore, by taking $\ell_{\rm ann}^{\rm InterG} \gg R$, we recover the estimate from \eref{JEG} upto the cross-section with $(\pi r_s^2)^{\rm TG}$ becoming $\int {\rm d}V(\eta_2(\hat{r})/r_s)^{\rm TG}$ in the $d \gg r_s$ limit. This variation in the cross-section is expected as all paths through the galaxy are not of equal thickness. For larger cross-sections, we calculate $\Phi_{\rm halo}$ numerically by $\Phi_{\rm halo} = d^2 \left(J_{\rm ann}-J_{\rm IG}\right)$ in the $d \gg r_s$ limit.

In \fref{GCE3}, we also present the $J$-factor ratios for the extra-galactic contribution. The contribution is comparable to the intra-galactic contribution, dominating slightly for Sextans and sub-dominant for Draco. The dip at $\sigma_2^{\rm b}/m_2\sim10^{5}~{\rm GeV}^{-3}$ is due to fewer $\chi_2^{\rm b}$ escaping from their source galaxy as observed by the simultaneous transition in the $(J/J^{\rm MW})_{\rm IG}$.

One effect we have not taken into account is the blocking of the external $\chi_2$ flux due to the presence of other galaxies. Although we take the escaping probability ${\rm exp}\left[-\Lambda\,\eta_2(\hat{r})\right]\sim 1$ in the small $\Lambda$ limit, this assumption can fail if $\chi_2$'s fly across many galaxies. To see this is not actually the case, we can calculate the solid angle in the sky that is occupied by Milky Way size galaxies (assuming core radius $r_s\sim 10$ kpc). A single galaxy that is $\hat{r}$ away from us covers a fraction of the sky $\sim\frac{\pi\,r_s^2}{4\pi \hat{r}^2}$. The total fraction of the sky being covered by all galaxies is
\begin{equation}
\int d\hat{r}\,4\pi\hat{r}^2\,n_{\rm halo}\,\frac{r_s^2}{4\,\hat{r}^2}\sim 0.1\,.
\end{equation}
This means $\chi_2$ produced in one galaxy only has a $10\%$ chance to hit another galaxy before reaching the target galaxy. Thus, when the escaping probability in each galaxy is close to one, this blocking does not change the $\chi_2$ flux significantly.

Besides creating the extra-galactic signal thus far discussed, these extra-galactic $\chi_2^{\rm b}$ can also annihilate with $\chi_2$ outside of galaxies and generate the isotropic gamma-ray background (IGRB) that is also measured by the Fermi-LAT experiment~\cite{Ackermann:2014usa}. This signal is produced by the inter-galactic annihilations discussed in the context of \eref{JEGfull} and can be simply derived by taking \eref{JEGfull} with the inter-galactic medium as the target. The generated IGRB flux
\begin{eqnarray}
&\frac{{\rm d}\Phi_{\gamma}^{\rm IGRB}}{{\rm d}E_\gamma{\rm d}\Omega}\sim\frac{n_{\rm halo}}{8\pi}\frac{\langle\sigma_1v\rangle}{m_1^2}\Phi_{\rm halo}R\left(1-e^{-R/\ell_{\rm ann}^{\rm InterG}} \right) \frac{{\rm d} N}{{\rm d}E_\gamma} \quad
\\
&\sim\frac{1.5 \times 10^{-9}}{{\rm cm^2}\,{\rm s}\,{\rm sr}}\left(\frac{\rho_{0}^2 f_1^2 \langle\sigma_1 v\rangle/m_1^2}{10^{-29}{\rm cm}^{-3}{\rm s}^{-1}}\right)\left(1-e^{-R/\ell_{\rm ann}^{\rm InterG}} \right) \frac{{\rm d} N}{{\rm d}E_\gamma},\nonumber
\end{eqnarray}
where $\rho_0$ is the characteristic density of MG, and we have assumed $r_s^{\rm MG}=20$~kpc. Interestingly, the result does not depend on $\chi_2$'s properties except in the exponential suppression, which will be minimal when this effect may be important ($\Lambda^{\rm MG} \ll 1$). The parameters used in \tref{param} produce a peak flux $E_\gamma^2{\rm d}\Phi_\gamma^{\rm IGRB}/{\rm d}E_\gamma{\rm d}\Omega \approx 3 \cdot 10^{-8}$ ${\rm GeV}\,{\rm cm^{-2}}\,{\rm s}^{-1}\,{\rm sr}^{-1}$. Comparing the flux with the IGRB bound $E_\gamma^2{\rm d}\Phi_{\gamma}^{\rm IGRB}/{\rm d}E_\gamma{\rm d}\Omega\lsim 10^{-7}~{\rm GeV}\,{\rm cm^{-2}}\,{\rm s}^{-1}\,{\rm sr}^{-1}$ derived in~\cite{Blanco:2018esa} for a similar gamma-ray spectrum, our signal should be well within the current constraint (especially once additional cosmological factors are taken into account as discussed above). Nevertheless, this diffuse gamma-ray background is a generic signature of the non-local annihilation model, and future experiments may be sensitive to it.

Additionally, producing $\chi_2^{\rm b}$ through $\chi_1$ annihilation in the inter-galactic medium is also feasible. However, since the average number density of DM particles in the inter-galactic medium is $\sim 10^{-5}$ smaller than in the galaxies, even if the volume of the observable universe is $\sim 10^{6}$ times larger than the sum of main galaxies, such $\chi_2^{\rm b}$ production is negligible.

\section{Reconciling the GCE as a signal of DM with dSph constraints} \label{sec.gce}

Since the non-local annihilation process suppresses the dSph gamma-ray signal relative to the signal from the MW comparative to canonical annihilation, an application of the non-local annihilation is to explain the potential mild tension between the DM explanation of the GCE signal~\cite{Hooper:2010mq} and the null-result in dSph observations (see e.g.,~\cite{Fermi-LAT:2016uux,Calore:2018sdx}). Note that while this discrepancy may not be very significant~\cite{Ando:2020yyk}, we discuss it here simply as an illustrative application of a specific NL model. The NL mechanism, however, is much broader and is independent of this particular result.

For canonical annihilation, a dSph signal produced by the same process as the GCE has been excluded to $\approx 2\sigma$.~\cite{Calore:2014xka,Ackermann:2015zua} The dSph signal needs to be suppressed by less than an order of magnitude in order to satisfy the bound. As we will show, the mild tension can be naturally addressed by the $\Lambda$ factor in NL annihilation. Additionally, the non-local signals with the distinct {\em fingerprints} in \fref{fingerprint} are only lower than the canonical annihilation signals by a factor of a few; they are thus still within the sensitivity of future observations.\footnote{Some other possible solutions to resolve this mild tension have been proposed in literature. For instance, in~\cite{Choquette:2016xsw}, the dSph signal is suppressed due to the $p$-wave DM annihilation process. In~\cite{Kaplinghat:2015gha}, the gamma-ray signal comes from the interaction between interstellar radiation and charged particles produced from the DM annihilation. These scenarios, however, predict much smaller dSph signals that are well below future observational sensitivity.}

In the right panel of \fref{GCE0}, the model labeled ``GCE", see \tref{param}, shows the required $\langle \sigma_{1} v\rangle$ for explaining the GCE via \erefs{step1}{step3}. We obtain the energy spectrum of the photons by numerically convolving the analytically calculated spectra of $\phi$ particles from the annihilation and the boosted spectra of an isotropic decay of $\phi$ into photons. Although the signal comes from a monochromatic decay, $\phi\to 2\gamma$, in $\phi$'s rest frame, since $\phi$ has a broad energy distribution from the $\chi_2^{\rm b}\chi_2$ annihilation, the ${\rm d}N_{\gamma}/{\rm d}E_{\gamma}$ also has a rather smooth spectrum.

We follow the technique outlined in Ref.~\cite{Calore:2014xka} for calculating the $\chi^2$ statistic for fitting to the GCE.\footnote{We use their covariance matrix with our predicted spectra
\begin{equation}
\chi^2=\sum_{ij} \left( \frac{{\rm d}\bar{N}}{{\rm d}E_i}(\boldsymbol{\theta})-\frac{{\rm d}N}{{\rm d}E_i}\right) \, \Sigma_{ij}^{-1}\, \left( \frac{{\rm d}\bar{N}}{{\rm d}E_j}(\boldsymbol{\theta})-\frac{{\rm d}N}{{\rm d}E_j}\right)
\end{equation}
where ${\rm d}N/{\rm d}E_i$ is the measured flux, ${\rm d}\bar{N}/{\rm d}E_i(\boldsymbol{\theta})$ is the predicted flux with input parameters $\boldsymbol{\theta}$, and $\Sigma_{ij}$ is the correlated covariance matrix.} As noted in Ref.~\cite{Calore:2014xka}, the best fit parameters may not visually appear to be optimal due to large cross-correlations between individual bins. The reduced $\chi^2$ for our model is 2.03 compared with 1.08 (1.52) for canonical $b\bar{b}$ ($\tau\bar{\tau}$) annihilation obtained in Ref.~\cite{Calore:2014xka} with 22 d.o.f. While the significance for this toy-model is weaker than other more standard models, it is is used here solely as an example of the behavior rather than a claim to fit the GCE. We obtain the best fit of the GCE signal with the spectrum as shown in \fref{GCE2} with $m_1=5.68~{\rm GeV}$. For comparison, we also show the best fit spectra for canonical $\chi\chi\to b\bar{b}$ and $\chi\chi\to \tau\bar{\tau}$ annihilation. The result has only a mild dependence on $m_{2,\,\phi,\,X}$ as long as $m_1\gg m_{2,\,\phi,\,X}$. For concreteness, we take $m_2=m_{\phi}=m_X=1~{\rm MeV}$ for the analysis. For a single set of model masses, the fitting routine has one additional free normalization parameter $f_0$, such that the observed flux from the GCE is 
\begin{equation}\label{eq.PhiGCE}
\frac{\Phi_{\rm GCE}}{\Delta\Omega_{\rm ROI}} = f_0 \int_{E_{\rm min}}^{E_{\rm max}}\frac{{\rm d}N\gamma}{{\rm d}E_\gamma}{\rm d}E_\gamma
\end{equation}
For each value of a set of $m_1$ with fixed $m_{2, \,\phi, \,X}$, we optimized $f_0$ to produce the minimum $\chi^2$. We then compared all the $\chi^2$s to find the global best fit $m_1$. Comparing with \eref{ndep1} and making proper conversions, it is obvious that for NL annihilation
\begin{equation}
f_{0,{\rm NL}} = \frac{\langle\sigma_1 v\rangle}{8\pi\,m_{1}^2}\times \frac{J_{\rm NL}}{\Delta\Omega_{\rm ROI}}
\label{eq.f0}
\end{equation}
$f_0$ is thus equivalent to the observed event rate per solid angle and is used to place constraints on $\langle\sigma_{1}v\rangle$ in \fref{GCE0}. We take the region-of-interest (ROI) to be $2^\circ<\theta<20^\circ$ from the galactic center where we have omitted the $\theta<2^\circ$ as in Ref.~\cite{Calore:2014xka}.

\begin{figure}
\includegraphics[width=0.95\columnwidth]{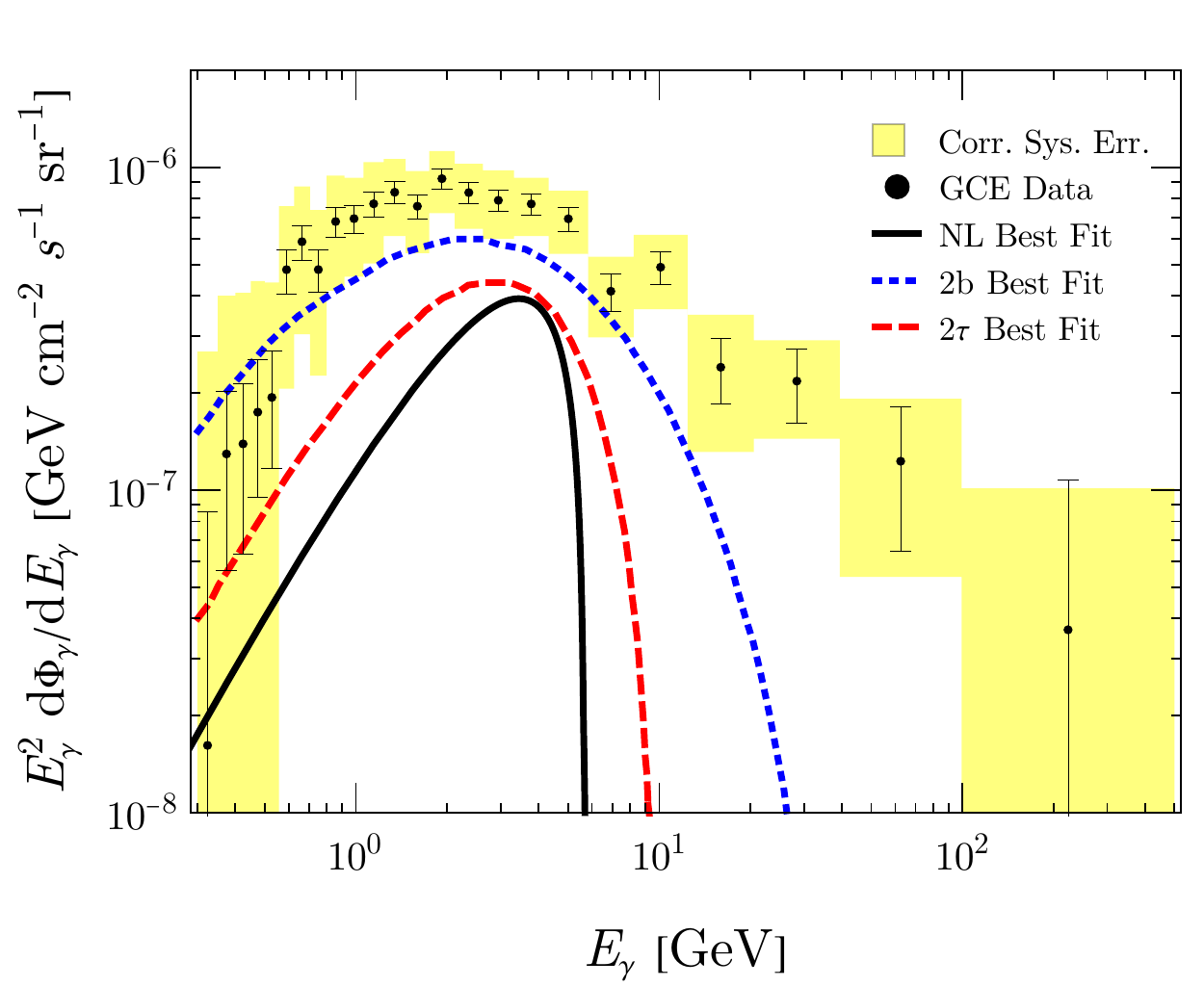}
\caption{Best fit gamma-ray spectrum for our toy-model $\chi_1\chi_1\to \chi_2^{\rm b}X, \; \chi_2^{\rm b}\chi_2\to 2\phi, \; \phi\to 2\gamma$. In the fit, most masses were fixed $m_2=m_\phi=m_X=1~{\rm MeV}$. These produced a best fit value of $m_1=5.68~{\rm GeV}$. Note that for $m_1\gg m_{2,/,\phi,\,X}$, the spectrum is largly independent of $m_{2,/,\phi,\,X}$. For comparison, the best fit for the canonical $b\bar{b}$ (dotted blue line) and $\tau\bar{\tau}$ annihilations (dashed red line) are also shown.}\label{fig.GCE2}
\end{figure}

As before, constraints on the photon-injection around recombination sets an upper bound on $\langle \sigma_{1} v\rangle$ via CMB measurements. Together with the bound from keeping $\alpha_2$ in \eref{crosssemi} perturbative, we find a window $1.7\times 10^3<\sigma_{2}^{\rm b}/m_2<3.3\times 10^3$~GeV$^{-3}$ for which NL $\chi_2^{\rm b}$ annihilations can produce a sufficient signal to explain the GCE. Based on \fref{GCE3}, we therefore expect a factor of $\approx 3\,(15)$ suppression for NL over canonical annihilation in the dSph/MW signal comparison. Thus, the suppression is enough to explain the absence of gamma-ray excess from the existing dSph observations, while suggesting that dSph signals can still be observed in the future.

\section{Conclusion}
In this paper, we have explored the scenario where the indirect detection signal comes from two consecutive DM annihilations. The boosted DM produced from the first annihilation can travel a long distance before annihilating with another at-rest DM particle into gamma-rays. This means the production of indirect detection signals becomes non-local with respect to the first annihilation. In fact, signals from a galaxy can arise either from boosted DM particle production and annihilation in the same galaxy (intra-galactic) or be triggered by boosted DM particles coming in from different, far away, galaxies (extra-galactic).

A robust consequence of the non-local annihilation is that the $J$-factor of the gamma-ray signal is different from those of the canonical DM annihilation and decay due to a further dependence on the DM density and size of the halo and an added dependence on the particle physics of the second annihilation. This implies that the associated ``ratio-of-ratios" (the ratio of the signal between two galaxies, say dSph vs. MW, as well as a comparison between the non-local and canonical models) will actually vary between galaxies. The non-local modification is thus galaxy-dependent. As we show in \fref{fingerprint}, if DM distributions in these dSphs follow the NFW distribution, we will be able to distinguish different DM scenarios once we see the gamma-ray signal from an ensemble of galaxies.

The magnitude of this effect on the ratio-of-ratios heavily depends on the second annihilation cross-section, see \fref{GCE3}. Indeed, in the extreme case of very large DM annihilation cross-section, requiring masses below $\mathcal{O}(10)$~MeV scale and/or couplings near the perturbative limit, the non-local model mimics the canonical scenario. Whereas, it is when the annihilation is less efficient that the $J$-factor ratio for the non-local model differs from the canonical model. However, in this opposite limit of much smaller annihilation cross-section, the ratio of $J$-factors will actually be independent of the annihilation cross section. In this case, the effect still carries additional galaxy dependence as compared to canonical annihilation. This is the case observed in \fref{fingerprint}. We thus obtain a prediction for the $J$-factor ratios for the non-local model based on only galactic parameters. It is important to point out that in the ``intermediate'' regime of annihilation cross-sections, the ratio of $J$-factors also depends on the cross-section, providing a means for measuring this annihilation rate.

The non-local annihilation process not only generates distinct galaxy-dependent signals, but can also reconcile the mild tension between gamma-ray signals from the MW and dSphs, namely explaining the DM annihilation interpretation of the GCE and the null result from dSph. The crucial observation is the gamma-ray signal from dSphs compared with the MW is smaller in the non-local scenario than it is for canonical annihilation, thus explaining the lack of dSphs gamma-ray signals in the current observation. However, unlike the explanation in~\cite{Kaplinghat:2015gha,Choquette:2016xsw}, the suppression of the dSph signal from the non-local process is only by a factor of a few and would be detectable with slight sensitivity improvements in dSph measurements.

Here we present some additional examples for future work \cite{NL_local_in_progress}. While in this work, we have discussed the signal using a specific asymmetric DM model with NFW profiles, there are many other scenarios that give the non-local annihilation as long as the DM sector produces boosted particles that have a large annihilation cross section with the ambient DM, but the same annihilation process has not been able to deplete the DM density. For example, the non-local annihilation process may also happen in a forbidden DM setup~\cite{Griest:1990kh,DAgnolo:2015ujb}. Similar to the model in \erefs{step1}{step3} but with $m_{\phi}>m_2$, the relic abundance of both $\chi_2$ and $\chi_2^*$ can be maintained due to the kinematic barrier even with a large $|\chi_2|^2\phi^2$ coupling. However, this barrier is overcome by the boosted DM. In this case, the non-local annihilation comes from $\chi_1\chi_1^*\to \chi_2^{\rm b}\chi_2^{{\rm b}*}$ and $\chi_2^{\rm b}\chi_2^*\to 2\phi(2\gamma)$. Moreover, although we assume the $X$ in \eref{step1}, which produces the boosted dark matter, to be an invisible particle for simplicity, $X$ can also be the $\phi$ particle that generates other gamma-ray signals at a different location from the $\chi_2^{\rm b}$ annihilation. This generates another interesting profile of the gamma-ray signal. Additionally, most of our discussion has focused on producing a gamma-ray signal; however, another source of comparison would be between neutrino production rates and the observed astrophysical neutrino flux~\cite{Aartsen:2020aqd}. As stated before, we concentrated in this work on the NFW profile. Qualitatively, other profiles, for example the cored Burkert profile~\cite{Burkert:1995yz}, exhibit the same NL features because they are due to $\chi_2^{\rm b}$ escaping from their parent galaxy. However, each profile's signal will possess different radial dependencies as well as a different allowed parameter space.

Finally, while a GCE explanation is intriguing and is certainly possible with DM mass of several GeV, in our model, in order to naturally generate a non-local signal, the boosted DM should have energy below $100$ MeV, and the resulting gamma-ray signal can be close to the threshold of the Fermi-LAT experiment. However, future proposals such as the e-ASTROGAM experiment are designed to cover the less-explored $1-100$ MeV gamma-ray region and can better probe non-local signals.

To conclude, the non-local framework is a natural outcome of multiple extended dark matter models and predicts additional galaxy dependencies in annihilation signals. This additional dependence results in smaller galaxies having an even smaller signal compared with larger counterparts. The comparison between the non-local behavior and the canonical framework for different galactic parameters stretches from a maximal difference to naturally merging with the canonical.

\section*{\uppercase{Acknowledgments}}

We would like to thank Patrick Fox, Roni Harnik, Dan Hooper, Rebecca Leane, Louis Strigari, and Yue Zhang for useful discussions. KA and YT were supported in part by the National Science Foundation under Grant Numbers PHY-1620074 and PHY-1914731, the Fermilab Distinguished Scholars Program, and the Maryland Center for Fundamental Physics. SJC was supported by the Brown Theoretical Physics Center. BD was supported in part by DOE Grant DE-SC0010813. YT was also supported in part by the US-Israeli BSF grant 2018236. YT thanks the Aspen Center for Physics, which is supported by National Science Foundation grant PHY-1607611, where part of this work was performed. KA and BD would like to thank the organizers of the ``2018 Santa Fe Summer Workshop in Particle Physics" where the ideas leading to this work were initiated.


\bibliographystyle{JHEP}
\bibliography{bibliography}


\appendix
\section{Alternate forms of the $J$-factor} \label{a.Jfacor_alt_derv}
The $J$-factors are written in multiple forms making use of various assumptions throughout the text; in this appendix, we derive these simplifications. The differential flux from an interaction seen by an observer is commonly written as \eref{ndep1}. It can also be compactly written for different interaction types as
\begin{equation}
\frac{{\rm d} \Phi_\phi}{{\rm d} E} = \frac{1}{4\pi} \int_{\rm ROI} {\rm d}\Omega_\ell \int_{\rm los} {\rm d}\ell\,\frac{{\rm d}N(r)}{{\rm d}V{\rm d}t}\frac{{\rm d}N_\phi}{{\rm d}E}
\label{eq.flux_basic}
\end{equation}
where $\phi$ is just a product from the interaction. The integral is taken over the line-of-site (los) and region-of-interest (ROI) observed. ${\rm d}N(r)/{\rm d}V{\rm d}t$ is the interaction rate per unit volume and time. ${\rm d}N_\phi/{\rm d}E$ is the spectrum of $\phi$ from the interaction. This form assumes that the interaction is spherically symmetric. As mentioned in the text, \eref{flux_basic} is typically separated into two parts, namely the astrophysical and the particle physics parameters. For canonical annihilating dark matter, \eref{flux_basic} can be written identically to \eref{ndep1} using
\begin{equation}
J_{\rm ann}=\frac{2 m_{\chi}^2}{\left\langle \sigma_{\rm ann} v \right\rangle} \int_{\rm ROI} {\rm d}\Omega_\ell \int_{\rm los} {\rm d}\ell\,\left( \frac{{\rm d}N(r)}{{\rm d}V{\rm d}t}\right)_{\rm ann}
\end{equation}
with
\begin{equation}
\left( \frac{{\rm d}N(r)}{{\rm d}V{\rm d}t}\right)_{\rm ann} = \frac{\left\langle \sigma_{\rm ann} v \right\rangle}{2 m_{\chi}^2} \rho_\chi^2(r)
\end{equation}
where $\chi$ is the dark matter particle with mass density $\rho_\chi$. A similar expression can be written for decay.  For a general expression, it is convenient to define
\begin{equation}
J=\int_{\rm ROI}{\rm d}\Omega_\ell \int_{\rm los} {\rm d}\ell \,f(r)\,,
\label{eq.J_full}
\end{equation}
where $f(r)$ is a scaled version of the number density of events per time which generates the signal, ${\rm d}N/{\rm d}V{\rm d}t$. This scaling is performed in such a way as to remove all possible particle physics contributions. We assume $f(r)$ is a spherically symmetric function centered at $\vec{\ell}=(d,0,0)$ in $\ell$ coordinate system. In the canonical case, $f(r)$ is $\rho_\chi^2(r)$ for annihilation and $\rho_\chi(r)$ for decay.

\subsection{$J$-factor in the $d\gg r_s$ limit}
The $J$-factors shown in \eref{J} assume the observer is far from the galaxy, $d\gg r_s$, such that all points in the galaxy can be treated as at equal distance. In order to demonstrate this far distance approximation, we restore some of the simplifications to the volume integral \eref{J_full} producing
\begin{equation}
J=4\pi\int \frac{{\rm d}V_\ell}{4\pi\,\ell^2} \,f(r)\,.
\label{eq.J_simp_1}
\end{equation}
For simplicity, we assume that we have captured all of the signal from the galaxy and have thus taken the integration over the volume of all space. ${\rm d}V_\ell$ indicates the integral is performed with $\ell$ coordinates. The $1/4\pi\,\ell^2$ is a result of an area suppression of flux with distance. The volumetric integral can easily be shifted to a new coordinate system centered at $r$ leading to
\begin{equation}
J=\int \frac{{\rm d}V_r}{\,\ell^2} \,f(r)\,.
\label{eq.J_simp_2}
\end{equation}
Finally, because the profiles have a cutoff scale $r_{\rm cutoff} \sim r_s$ and $d\gg r_s$, the integral is dominated by the region where $r\ll d$ and thus $|\vec{\ell}| = |\vec{d} - \vec{r}| \approx |\vec{d}|$. This results in the simplification quoted in \eref{J}. Note that a cutoff scale must be imposed for decay with an NFW profile because at arbitrarily large distances, its volumetric integral is logarithmically divergent. In this work, we achieve this through our choice of boundaries in our ROI and los integrations. Tests showed that different approaches resulted in differences at the percent level. We can also write in this limit the $J$-factors in the dimensionless integral format as defined in this work
\bea
J_{\rm ann} & = & \frac{\rho_0^2\,r_s^3}{d^{2}}\int {\rm d^3}\hat{r}\, \eta^2(\hat{r}), 
\label{eq.J_ann_dimensionless} \\
J_{\rm dec} & = & \frac{\rho_0\,r_s^3}{d^{2}}\int {\rm d^3}\hat{r} \, \eta(\hat{r})\,.
\label{eq.J_dec_dimensionless}
\eea

\section{$J_{\rm NL}$ derivation} \label{a.J_NL_derivation}
In this appendix, we derive the $J$-factors and associated functions that arise from the non-local annihilation framework, primarily focusing on models where the boosted DM is produced via another annihilation within the same galaxy. In a model where the observed dark matter signal is produced through a secondary interaction, the two interaction events do not necessarily need to occur at the same location in space. Let us consider a two-component dark matter annihilation model with particles $\chi_1$ and $\chi_2$. The annihilation of $\chi_1$ produces a boosted $\chi_2$ referred to from here on as $\chi_2^{\rm b}$. Due to the current conditions, $\chi_2$ is unable to annihilate with itself, but it can annihilate with $\chi_2^{\rm b}$. The general model setup is that one set of dark matter, $\chi_1$, annihilates into another variety, $\chi_2$, but with a non zero velocity, $\chi_2^{\rm b}$. The boosting allows it to access otherwise forbidden channels; depending on the cross-section, $\chi_2^{\rm b}$ may annihilate at a different location from its creation.

For the discussion that follows, we assume all annihilations occur within a galaxy. Subscripts 1 and 2 correspond to the various parameters for particles $\chi_1$ and $\chi_2$, respectively.

\subsection{$\chi_2^{\rm b}$ survival probability}

Before calculating the spatial distribution of the $\chi_2^{\rm b}\chi_2$ annihilation, let us derive the probability function ${\rm d}\mathcal{P}(r,s)/{\rm d}s$ of having a $\chi_2^{\rm b}$ being produced at $s=0$ (from the $\chi_1$ annihilation) and annihilating at a distance $s$ away, see \fref{coordinate_appendix}.

\begin{figure}
\begin{center}
\includegraphics[width=0.95\columnwidth]{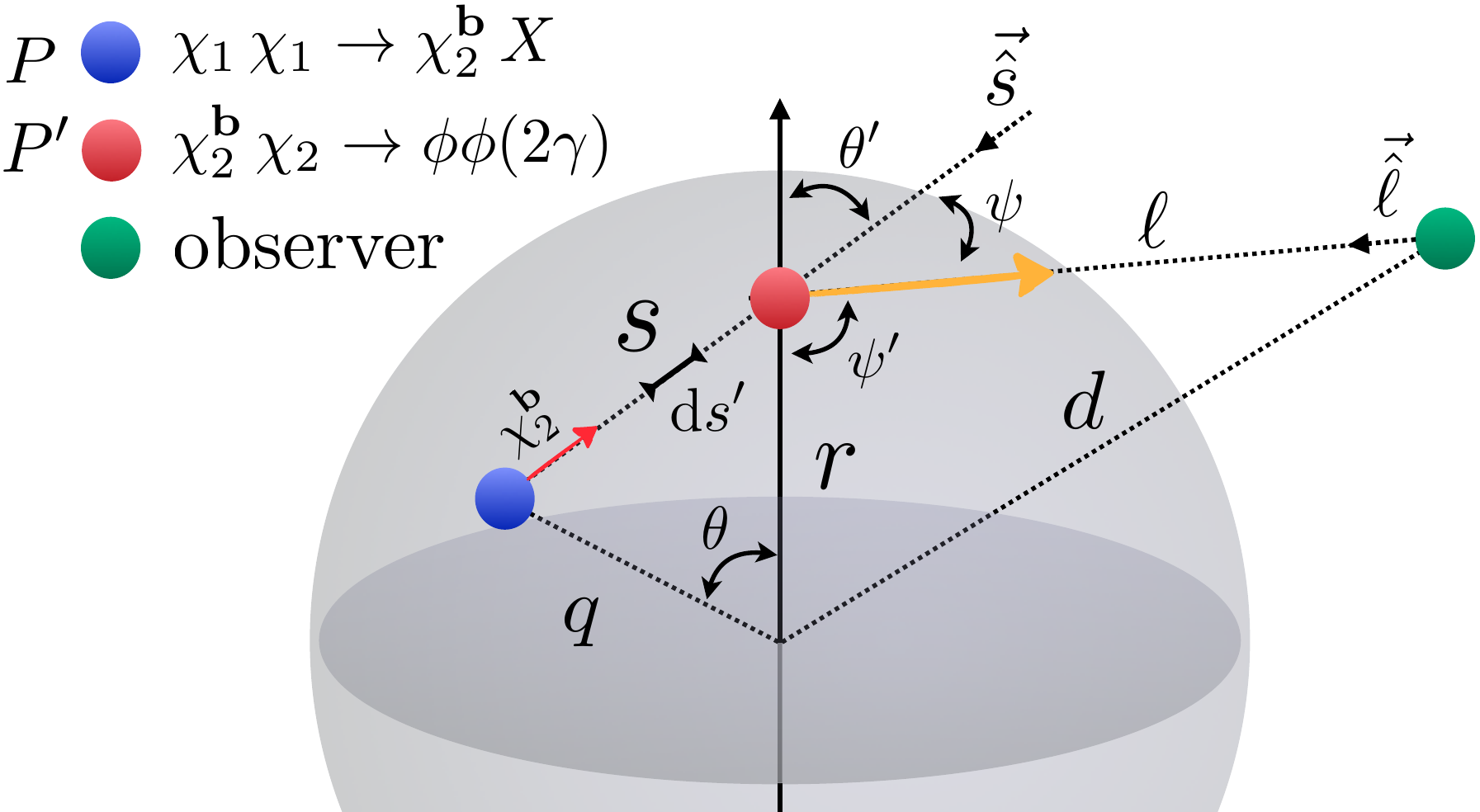}
\caption{A modified version of \fref{coordinate} to highlight particular integration angles. A $\chi_1\chi_1$ annihilation first occurs at the blue point $P$ a distant $q$ from the halo's center. The produced $\chi_2^{\rm b}$ travels a distance $s$ and annihilates with a slow moving ambient $\chi_2$ at the red point $P^\prime$ into $\phi$'s that decay promptly on galactic scales into gamma-rays which are observed at the green point. $\phi$'s are produced isotropically in the $\chi_2^{\rm b}\chi$ rest frame, but are boosted in the observer's reference frame. This introduces an angular spectrum that is dependent on $\psi$, the angle between $\vec{\hat{\ell}}$ and $\vec{\hat{s}}$}
\label{fig.coordinate_appendix}
\end{center}
\end{figure}

Let us first assume the number density of $\chi_2$ is constant, $n_2(s)=n_2$. When slicing the distance $s$ into infinitesimally small pieces of length $\Delta s$, the probability of having $\chi_2^{\rm b}$ to not have annihilated after traveling $s$ but annihilated before $s+\Delta s$ can be written as
\begin{eqnarray}
\Delta \mathcal{P}(s) &=& \left(1-n_2 \sigma_2^{\rm b} \Delta s\right)^{\frac{s}{\Delta s}}\left(n_2\sigma_2^{\rm b} \Delta s\right),
\nonumber \\
&=& \exp\left[\frac{s}{\Delta s}\ln\left(1-n_2 \sigma_2^{\rm b} \Delta s\right)\right]\left(n_2 \sigma_2^{\rm b} \Delta s\right)
\nonumber  \\
&\approx&e^{-n_2 \sigma_2^{\rm b}\,s}n_2\sigma_2^{\rm b} \Delta s,
\end{eqnarray}
where $\sigma_2^{\rm b}$ is the cross-section for a $\chi_2^{\rm b} \chi_2$ annihilation. Here we assume fine enough divisions on $s$ such that the probability of having annihilations in each $\Delta s$ window $n_2 \sigma_2^{\rm b} \Delta s\ll 1$. This gives the probability function 
\begin{equation}
\frac{{\rm d}\mathcal{P}(s)}{{\rm d}s}=e^{-n_2 \sigma_2^{\rm b} \,s}\,n_2 \sigma_2^{\rm b}
\end{equation}
for a constant $\chi_2$ density. When $n_2 \sigma_2^{\rm b} s>1$, the chance for $\chi_2^{\rm b}$ to survive is exponentially suppressed as a function of distance. When $n_2 \sigma_2^{\rm b} s\ll 1$, $\chi_2^{\rm b}$ is unlikely to have annihilated, and the chance of annihilating in a short distance is a constant ($n_2 \sigma_2^{\rm b}$) as expected.

If $n_2(s)$ is instead a smoothly varying function of distance, we can again divide the distance into infinitesimal $\Delta s$ pieces, such that the $n_2(s_i)$ in each $[s_i, s_i+\Delta s]$ piece is almost a constant. In this case, the probability of $\chi_2^{\rm b}$ annihilating in-between  $s$ and $s+\Delta s$ can be written as ($s_0\equiv0$)
\begin{eqnarray}
\frac{{\rm d}\mathcal{P}(s)}{{\rm d}s}&=&\displaystyle{\prod_{i=0}^{s/\Delta s}}\left(1-\int_{0}^{\Delta s}{\rm d}\hat{s}\,e^{-\bar{n}_i\sigma_2^{\rm b} \hat{s}}\sigma_2 \bar{n}_i\right)\,n_2(s)\,\sigma_2^{\rm b},
\nonumber \\
&=&\displaystyle{\prod_{i=0}^{s/\Delta s}}\left(1+e^{-\bar{n}_i\sigma_2^{\rm b} \Delta s}-1\right)\,n_2(s)\,\sigma_2^{\rm b},
\nonumber \\
&=&\exp\left[-\displaystyle{\sum}_{i=0}^{s/\Delta s}\bar{n}_i\sigma_2^{\rm b} \Delta s\right]\,n_2(s)\,\sigma_2^{\rm b}.
\end{eqnarray}
Taking the limit $\Delta s/s \to 0$, we have the probability function for a general $n_2(r)$
\begin{equation}
\frac{{\rm d}\mathcal{P}(r,s)}{{\rm d}s}=\exp\left[-\int_{0}^{s}{\rm d}\tilde{s}\,n(\tilde{s})\sigma_2^{\rm b} \right]\,n_2(r)\,\sigma_2^{\rm b}.
\label{eq.dP_ds}
\end{equation}
where we have further generalized ${\rm d}\mathcal{P}/{\rm d}s$ to be the probability to annihilate at point $r$ after traveling a distance $s$. \eref{dP_ds} is \eref{prob} which appeared in the main text with a few cosmetic alterations.

\subsection{Rate of the secondary annihilation}

Here we derive the $\chi_2^{\rm b} \chi_2$ annihilation rate per volume as a function of radius $r$ from the halo center denoted by
\begin{equation}
\frac{{\rm d}N_2(r)}{{\rm d}V\,{\rm dt}}
\end{equation}
We define the coordinates as in \fref{coordinate_appendix}, where we assume a spherically symmetric halo density profile $n_2(r)$ and want to calculate the $\chi_2^{\rm b} \chi_2$ annihilation rate at the red dot $P^\prime$. Since the result will only depend on $r$, we can put the red point on the $z$-axis and integrate over the $\chi_2^{\rm b}$ coming from $\chi_1$ interactions at each blue point $P$ around the halo (i.e., integrating over $(s,\theta^\prime)$) to obtain the total $\chi_2^{\rm b} \chi_2$ annihilation rate. Note that the center of integration is taken at the second annihilation location rather than the center of the halo. This choice is to aid in the inclusion of a non-spherically symmetric annihilation distribution for the second annihilation originating from the boosted particle's trajectory.

First, $\chi_1$ annihilations happen at point $P$ (blue) and produce $\chi_2^{\rm b}$. This is followed by $\chi_2^{\rm b} \chi_2$ annihilation at point $P^\prime$ (red) with the rate given by
\bea
\frac{{\rm d}N_2(r)}{{\rm dt}} & = & \int{\rm d}^3\vec{s}\,\frac{{\rm d}n_1(q)}{{\rm d}t}\,\frac{\Delta A_s}{4\pi\,s^2}\,\mathcal{P}(r,s) \nonumber \\
& = & \int {\rm d}^3\vec{s}\,\frac{{\rm d}n_1(q)}{{\rm d}t}\,\frac{1}{4\pi\,s^2}\, \frac{{\rm d}\mathcal{P}(r,s)}{{\rm d}s}\Delta V_s.
\eea
Here $n_2(q)$ is the number of $\chi_1$ annihilation per volume at radius $q$, and $\mathcal{P}(r,s)$ is the probability of $\chi_2^{\rm b}$ annihilating after being produced from the blue point and then traveling to the red point. This probability is derived in the previous section. We assume the $\chi_2^{\rm b}$ are produced isotropically from the $\chi_1$ annihilation, and the probability of having $\chi_2^{\rm b}$ reach the red point is suppressed by dilution of the flux with distance, $\Delta A_s/4\pi s^2$, where $\Delta A_s$ is the infinitesimally small area of the red point. After plugging in \eref{dP_ds}, the rate density can be written as
\begin{equation}
\frac{{\rm d}N_2(r)}{{\rm d}V{\rm d}t}=\int\frac{{\rm d^3}\vec{s}}{4\pi s^2}\,\frac{{\rm d}n_1(q)}{{\rm d}t}\times \frac{{\rm d}\mathcal{P}(r,s)}{{\rm d}s},
\end{equation}
where
\begin{eqnarray}
\frac{{\rm d}n_1(q)}{{\rm d}t} &=& \frac{(\rho_1(q))^2\langle\sigma_1\,v\rangle}{2m_1^2}\\
q = q(r,s,\cos\theta^\prime)&=&\sqrt{r^2+s^2-2rs\cos{\theta^\prime}},
\end{eqnarray}
and the probability function
\bea
\frac{{\rm d}\mathcal{P}(r,s)}{{\rm d}s}=\exp\left[-\int_0^{s}{\rm d}\tilde{s}\,n_2(\tilde{s})\sigma_2 \right]\sigma_2 \,n_2(r),\\
n_2(\tilde{s})=n_2\,\,{\rm at\,\,radius}\,\,\sqrt{r^2+\tilde{s}^2-2r\tilde{s}\cos \theta^\prime}.
\eea

By defining dimensionless lengths, $\hat{r}=r\,r_s$, we can further simplify these expressions down to normalized density distributions, $\eta_i(\hat{r})=n_i(r_s\,\hat{r})/n_{i,0}$, and a single scale factor, $\Lambda=n_{2,0} \sigma_2 r_s$.
\bea
\frac{{\rm d}N_2(r)}{{\rm d}V{\rm d}t} & = & \frac{n_{1,0}^2 \langle\sigma_1\,v\rangle}{2} \frac{\Lambda\,\eta_2(\hat{r})}{4\pi}\int\frac{{\rm d^3}\vec{\hat{s}}}{\hat{s}^2}\,(\eta_1(\hat{q}))^2 \nonumber \\
& & \quad \times \exp\left[-\Lambda \int_0^s{\rm d}\hat{s}\,\eta_2(\hat{s})\right].
\label{eq.dN2_dVdt}
\eea

Because we are working with boosted particles, we also define the angular annihilation density to preserve the particle velocities
\bea
\frac{{\rm d}N_2(r)}{{\rm d}V{\rm d}t\,{\rm d}\Omega_{\vec{s}}}& = & \int\frac{{\rm d}\vec{s}}{4\pi}\,\frac{{\rm d}n_1(q)}{{\rm d}t}\times \frac{{\rm d}\mathcal{P}(r,s)}{{\rm d}s} \nonumber \\
& = & \frac{n_{1,0}^2 \langle\sigma_1\,v\rangle}{2} \frac{\Lambda\,\eta_2(\hat{r})}{4\pi}\int {\rm d}\hat{s}\,(\eta_1(\hat{q}))^2 \nonumber \\
& & \quad \times \exp\left[-\Lambda \int_0^s{\rm d}\hat{s}\,\eta_2(q\hat{s})\right],
\label{eq.NL_angular_density}
\eea
where $\Omega_{\vec{s}}$ denotes the angular dependence. Note that due to spherical symmetry, the azimuthal integral is trivial, only appearing in ${\rm d}\Omega_{\vec{s}}$. However, due to a dependence in the signal from the angle between $\vec{s}$ and $\vec{\ell}$, it is left unintegrated here. This additional dependence is due to the introduction of the observer which breaks the spherical symmetry assumed up to this point.

By utilizing \eref{flux_basic}, \eref{J_full}, the annihilation rate from \eref{dN2_dVdt}, and also assuming the spectra from the second annihilation is isotropic, the differential flux for non-local annihilation is
\begin{equation}
\left(\frac{{\rm d} \Phi_\phi}{{\rm d} E}\right)_{\rm iso}=\frac{\left\langle \sigma_1 v \right\rangle}{8\pi m_1^2} \left( \frac{{\rm d}N_\phi}{{\rm d}E} \right) J_{\rm iso}
\label{eq.flux_isotropic}
\end{equation}
with
\begin{equation}
J_{\rm iso}=\frac{2m_1^2}{\left\langle \sigma_1 v \right\rangle}\int_{\rm ROI} {\rm d}\Omega_\ell \int_{\rm los} {\rm d}\ell\,  \frac{{\rm d}N_2(r)}{{\rm d}V{\rm d}t},
\label{eq.J_isotropic}
\end{equation}
similar to annihilation in \eref{ndep1} but with a different $J$-factor. Note that this formulation does not separate the astrophysics from all of the particle properties. It only removes the $\chi_1$ dependencies but leaves $\chi_2$ in the form of $\Lambda$, see \eref{Lambda}. When the boosted spectra is not isotropic, the differential flux is
\begin{equation}
\frac{{\rm d} \Phi_\phi}{{\rm d} E}=\frac{\left\langle \sigma v \right\rangle_1}{8\pi m_1^2} J
\label{eq.flux_exact_angular}
\end{equation}
with
\bea
J & = & \frac{8\pi m_1^2}{\left\langle \sigma v \right\rangle_1} \int_{\rm ROI} {\rm d}\Omega_\ell \int_{\rm los} {\rm d}\ell \int {\rm d}\Omega_{\vec{s}} \nonumber \\
& & \quad \times \, \frac{{\rm d}N_2(r)}{{\rm d}V{\rm d}t\,{\rm d}\Omega_{\vec{s}}} \frac{{\rm d}N_\phi}{{\rm d}E{\rm d}\Omega_{\vec{s}}}(\vec{s},\vec{\ell}) \label{eq.J_exact_angular}
\eea
where ${\rm d}N_\phi/{\rm d}E{\rm d}\Omega_{\vec{s}}(\vec{s},\vec{\ell})$ is the differential angular spectrum of the annihilation. ${\rm d}N_\phi/{\rm d}E{\rm d}\Omega_{\vec{s}}(\vec{s},\vec{\ell})$ depends on the angle between the $\chi_2^{\rm b}$'s direction of motion and the direction to the observer. Using the coordinates as shown in \fref{coordinate_appendix}, this angle is defined by $\cos\left(\psi\right)=\vec{\ell}\cdot\vec{s}/|\ell||s|$. Note that in order to keep the same leading factor in \eref{flux_exact_angular} and a normalized definition for the differential angular spectrum, an extra factor of $4\pi$ has been included in \eref{J_exact_angular}. This is because the normalization of ${\rm d}N_\phi/{\rm d}E{\rm d}\Omega_{\vec{s}}$ has already been included in \eref{NL_angular_density}. This factor can be easily identified for a uniform distribution where ${\rm d}N_\phi/{\rm d}E{\rm d}\Omega_{\vec{s}}=1/4\pi \times {\rm d}N_\phi/{\rm d}E$. Combining \eref{NL_angular_density} and \eref{J_exact_angular} yields the full intra-galactic $J$-factor defined in the text, \eref{J_nl_exact}:
\bea
J_{\rm IG} & = &\int_{\rm ROI}\, {\rm d}\Omega_\ell \int_{\rm los} {\rm d} \ell \int_{\vec{s}} \frac{{\rm d}^3\vec{\hat{s}}}{2\pi \hat{s}^2} \\
& & \times \frac{{\rm d}{\mathcal P}_{\chi_2^{\rm b}\chi_2}(\hat{r},\hat{s})}{{\rm d}\hat{s}}\,[\rho_{1,0}\eta_1(\hat{q})]^2 \frac{{\rm d}N}{{\rm d}E{\rm d}\Omega_{\vec{s}}}(\vec{\hat{s}},\vec{\ell})\,\nonumber
\eea
Note that this version is more general than \eref{J_nl_exact} as explained below. Due to anisotropies, the spectral dependencies of the interaction are not separable from the rest of the calculation. But, in the highly boosted case, we assume ${\rm d}N_\phi/{\rm d}\Omega_{\vec{s}}\propto\delta(\psi)$ and the distribution becomes separable
\begin{equation}
\frac{{\rm d}N_\phi}{{\rm d}E{\rm d}\Omega_{\vec{s}}}=\frac{{\rm d}N_\phi}{{\rm d}E}\frac{{\rm d}N_\phi}{{\rm d}\Omega_{\vec{s}}}=\frac{1}{4\pi}\frac{{\rm d}N_\phi}{{\rm d}E}\frac{{\rm d}N_\phi}{{\rm d}(\cos(\theta^\prime))},
\label{eq.angular_dep}
\end{equation}
where ${\rm d}N_\phi/{\rm d}(\cos(\theta^\prime))=\delta(\psi)$, as all of the spectrum is highly peaked in the direction of $\chi_2^{\rm b}$'s momentum. In order to match the form in \eref{ndep1}, \eref{J_nl_exact} is written with this approximation that the energy spectrum is separable from the angular spectrum.

As noted by \eref{angular_dep}, in this delta function limit, the azimuthal dependence is trivial and the zenith angle is restricted to $\psi=\pi-\theta^\prime-\psi^\prime=0$ with $\cos\left(\psi^\prime\right)=(r^2+\ell^2-d^2)/2\,r\,\ell$, thus $\cos(\theta^\prime)=-\cos(\psi^\prime)$ and ${\rm d}N_\phi/{\rm d}(\cos(\theta^\prime))=\delta(\cos(\theta^\prime)+\cos(\psi^\prime))$. These angular dependencies in the delta function limit permit the trivialization of the ${\rm d}\Omega_{\vec{s}}$ integration, leaving just the ${\rm d}s$ integral. The final result only depends on the distributions $\eta_i$ and $\Lambda$, as observed in \erefs{dN2_dVdt}{NL_angular_density}, and the observer integrations over ROI and los.

\end{document}